\documentclass[aps,pre,showpacs,floatfix]{revtex4}
\usepackage{epsf}
\usepackage{amsmath}
\usepackage{amssymb}
\usepackage{graphicx}
\def\mathbi#1{\textbf{\em #1}}
\newcommand{\be}{\begin{equation}}
\newcommand{\ee}{\end{equation}}
\newcommand{\bea}{\begin{eqnarray}}
\newcommand{\eea}{\end{eqnarray}}

\begin{document}

\title{\bf Thermodynamic time asymmetry in nonequilibrium fluctuations}

\author{D. Andrieux and P. Gaspard}
\affiliation{Center for Nonlinear Phenomena and Complex Systems,\\
Universit\'e Libre de Bruxelles, Code Postal 231, Campus Plaine,
B-1050 Brussels, Belgium}

\author{S. Ciliberto, N. Garnier, S.
Joubaud, and A. Petrosyan}
\affiliation{Laboratoire de Physique, CNRS UMR 5672, Ecole Normale
Sup\'erieure de Lyon,
46 All\'ee d'Italie, 69364 Lyon C\'edex 07, France}

\begin{abstract}
We here present the complete analysis of experiments on driven Brownian motion
and electric noise in a $RC$ circuit, showing that thermodynamic
entropy production
can be related to the breaking of time-reversal symmetry in the
statistical description
of these nonequilibrium systems.  The symmetry breaking can be
expressed in terms of dynamical
entropies per unit time, one for the forward process and the other
for the time-reversed
process.  These entropies per unit time characterize dynamical
randomness, i.e., temporal
disorder, in time series of the nonequilibrium fluctuations.  Their
difference gives the
well-known thermodynamic entropy production, which thus finds its origin in the
time asymmetry of dynamical randomness, alias temporal disorder, in systems
driven out of equilibrium.
\end{abstract}


\maketitle

\section{Introduction}

According to the second law of thermodynamics, nonequilibrium
systems produce entropy in a time asymmetric way.  This
thermodynamic time asymmetry is usually expressed in terms of
macroscopic concepts such as entropy.  The lack of
understanding of this time asymmetry in terms of concepts closer
to the microscopic description of the motion of particles has
always been a difficulty. Only recently, general relationships
have been discovered which allows us to express the thermodynamic
time asymmetry at the mesoscopic level of description in terms of
the probabilities ruling the molecular or thermal fluctuations in
nonequilibrium systems. More specifically the entropy production
rate in a system driven out of equilibrium can be estimated by
measuring the asymmetries between  the probabilities of finding
certain time evolutions  when the system is driven with forward
nonequilibrium driving and those of finding the corresponding reversed time
evolutions with a backward driving. In a recent letter
\cite{AGCGJP07}, experimental evidence has been reported that,
indeed, the entropy production finds its origin in the time
asymmetry of the dynamical randomness, alias temporal disorder, in
driven Brownian motion and in the electric noise of a driven $RC$
circuit. In these two experiments we record long time series,
either of the position of the Brownian particle or of the
fluctuating voltage of the $RC$ circuits, which allow us to define
the probabilities of given time evolutions, also called paths. This result
shows that, under nonequilibrium conditions, the probabilities of
the direct and time-reversed paths break the time-reversal symmetry
and that the entropy production is given by the difference between
the decay rates of these probabilities.
These decay rates characterize the dynamical
randomness of the paths (or their time reversals) so that the
entropy production turns out to be directly related to the
breaking of the time-reversal symmetry in the dynamical randomness
of the nonequilibrium fluctuations.

The purpose of the present paper is to provide a detailed report of
these two  experiments and of the data analysis. The dynamical
randomness of the forward time series is characterized by the
so-called  $(\varepsilon,\tau)$-entropy per unit time,
which represents the average decay rate of the probabilities
of paths sampled with a resolution
$\varepsilon$ and a sampling time $\tau$ \cite{GW93}.
The precise definition of these quantities will be given in Sec. \ref{NA}.
To each possible path in the forward time series, we look for the
corresponding time-reversed path in the backward time series. This
allows us to further obtain the time-reversed
$(\varepsilon,\tau)$-entropy per unit time.   Remarkably, the difference
between the backward and forward  $(\varepsilon,\tau)$-entropies
per unit time gives the right positive value of the thermodynamic entropy production
under nonequilibrium conditions. This result shows by direct data
analysis that the entropy production of nonequilibrium thermodynamics can be
explained as a time asymmetry in the temporal disorder
characterized by these new quantities which are the
$(\varepsilon,\tau)$-entropies per unit time.

These quantities were introduced in order to generalize the
concept of Kolmogorov-Sinai entropy per unit time from dynamical
systems to stochastic processes \cite{GW93,ER85}.  In this regard,
the relationship here described belongs to the same family of
newly discovered large-deviation properties as the escape-rate and
chaos-transport formulas \cite{GN90,DG95,GCGD01}, the steady-state
or transient fluctuation theorems
\cite{ECM93,ES94,GC95,K98,LS99,M99,MN03,W02,ZC03,ZCC04,GC04,DJGPC06,SSTWS05,S05,TC06,HS07,AG07a,AG07b},
and the nonequilibrium work fluctuation theorem
\cite{C99,DCP05,CRJSTB05,J06,KPV07}. All these relationships share
the common mathematical structure that they give an irreversible
property as the difference between two decay rates of mesoscopic
or microscopic properties \cite{G05,G06}.  These relationships are
at the basis of the important advances in nonequilibrium
statistical mechanics during the last two decades \cite{DGV07}.
Recently, the concept of time-reversed entropy per unit time was
introduced \cite{G04} and used to express the thermodynamic
entropy production in terms of the difference between it and the
standard (Kolmogorov-Sinai) entropy per unit time
\cite{G05,G04,LAvW05,NVdS06}. This relationship can be applied to
probe the time asymmetry of time series \cite{AGCGJP07,PRD07} and
allows us to understand the origin of the thermodynamic time
asymmetry.

The paper is organized as follows. In Sec. \ref{theo}, we introduce
the Langevin description of the experiments.
In particular, we show that the thermodynamic entropy production
arises from the breaking of the forward and time-reversed probability
distributions over the trajectories.
In Sec. \ref{NA}, we introduce the dynamical entropies and describe
an algorithm for their estimation using long time series. It is shown
that the difference between the two dynamical entropies gives back
the thermodynamic entropy production.
The experimental results on driven Brownian motion are presented in
Sec. \ref{BM} where we analyze in detail the behavior of the
dynamical entropies and establish the connection with the
thermodynamic entropy production.  The experimental results on
electric noise in the driven $RC$ circuit are given in Sec. \ref{RC}.
The relation to the extended fluctuation theorem
\cite{ZC03,ZCC04,GC04} is discussed in Sec. \ref{discussion}. The
conclusions are drawn in Sec. \ref{conclusions}.


\section{Stochastic description, path probabilities, and entropy production}
\label{theo}

We consider a Brownian particle in a fixed optical trap and
surrounded by a fluid moving at the speed $u$.
In a viscous fluid such as water solution at
room temperature and pressure, the motion of a dragged micrometric
particle is overdamped. In this case, its Brownian motion can be modeled by the following Langevin equation \cite{ZCC04}:
\begin{equation}
\alpha \frac{dz}{dt} = F(z)+\alpha u + \xi_t
\label{zm}
\end{equation}
where $\alpha$ is the viscous friction coefficient,
$F=-\partial_zV$ the force exerted by the potential
$V$ of the laser trap, $\alpha u$ is the drag force of the fluid moving at speed $u$,
and $\xi_t$ a Gaussian white noise with its average and correlation
function given by
\begin{eqnarray}
\langle \xi_t\rangle &=& 0 \\
\langle \xi_t\, \xi_{t'}\rangle &=& 2 \, k_{\rm B} T \, \alpha\, \delta(t-t')
\end{eqnarray}

In the special case where the potential is harmonic of stiffness $k$,
$V=kz^2/2$,
the stationary probability density is Gaussian
\begin{equation}
p_{\rm st} (z) = \sqrt{\frac{\beta k }{2 \pi }} \exp \Big[ -
\frac{\beta k}{2}(z-u\, \tau_R)^{2} \Big]
\label{Pst}
\end{equation}
with the relaxation time
\begin{equation}
\tau_R=\frac{\alpha}{k}
\label{t_R}
\end{equation}
and the inverse temperature $\beta=(k_{\rm B}T)^{-1}$. The maximum
of this Gaussian distribution is located at the distance $u\tau_R$
of the minimum of the confining potential.  This shift is due to
dragging and corresponds to the position where there is a balance
between the frictional and harmonic forces.

The work $W_t$ done on the system by the moving fluid during the time interval
$t$ is given by \cite{ZC03,ZCC04,S97}
\begin{equation}
W_t = -\int_0^t u \; F(z_{t'}) \; dt'
\label{work}
\end{equation}
while the heat $Q_t$ generated by dissipation is
\begin{equation}
Q_t =  \int_0^t \left(\dot{z}_{t'}-u\right) \; F(z_{t'}) \; dt'
\label{heat}
\end{equation}
We notice that the quantities (\ref{work}) and (\ref{heat})
are fluctuating because of the Brownian motion of the particle.
Both quantities are related by the change in potential energy $\Delta
V_t \equiv
V(z_t)-V(z_0)$ so that
\begin{equation}
Q_t = W_t - \Delta V_t
\end{equation}

In a stationary state, the mean value of the dissipation rate is equal to the
mean power done by the moving fluid
since $\lim_{t\to\infty}(1/t)\langle \Delta V_t\rangle=0$.
The thermodynamic entropy production is thus given by
\begin{equation}
  \frac{d_{\rm i}S}{dt} = \lim_{t\to\infty}
\frac{1}{t} \frac{\langle Q_t \rangle}{T}= \lim_{t\to\infty}
\frac{1}{t} \frac{\langle W_t \rangle}{T} = \frac{\alpha u^2}{T}
\label{mean}
\end{equation}
in the stationary state. The equilibrium state is reached
when the speed of the fluid is zero, $u=0$,
in which case the entropy production (\ref{mean})
vanishes as expected.

An equivalent system is an $RC$ electric circuit
driven out of equilibrium by a current source which imposes the mean
current $I$ \cite{ZCC04,GC04}. The current fluctuates in the circuit
because of the
intrinsic Nyquist thermal noise.  This electric circuit
and the dragged Brownian particle, although physically different,
are known to be formally equivalent by the correspondence shown in
Table \ref{table} \cite{ZCC04}.

\begin{table}[t]
\centerline{
\begin{tabular}{|c|c|}
\hline
Brownian particle & $RC$ circuit\\\hline
   $z_t$ & $q_t-It$ \\
   $\dot z_t$ & $\dot q_t-I$ \\
   $\xi_t$&$-\delta V_t$ \\
   $\alpha$  & $R$\\
   $k$ & $1/C$\\
   $u$  & $-I$\\
\hline
\end{tabular}
}
\caption{The analogy between the Brownian particle and the electric
$RC$ circuit. For the Brownian particle, $z_t$ is its position,
$\dot z_t$ its velocity, $\xi_t$ the Langevin fluctuating force,
$\alpha$ the viscous friction coefficient,
$k$ the harmonic strength or stiffness of the optical trap,
and $u$ the fluid speed.  For the electric circuit,
$q_t$ is the electric charge passing through the
resistor during time $t$, $i_t=\dot q_t$ the corresponding current,
$\delta V_t$ the fluctuating electric potential of the Nyquist noise,
$R$ the resistance, $C$ the capacitance,
and $I$ the mean current source.}\label{table}
\end{table}

Our aim is to show that one can extract the heat dissipated
along a fluctuating path by comparing
the probability of this path, with the probability of the
corresponding time-reversed path having
also reversed the external driving, i.e., $u \rightarrow -u$
for the dragged Brownian particle (respectively, $I\to -I$ for the $RC$ circuit).

We use a path integral formulation.
A stochastic trajectory is uniquely defined by specifying the noise
history of the system $\xi_t$.
Indeed, the solution of the stochastic equation (\ref{zm}), i.e.,
\begin{eqnarray}
z_t &=& z_0 + \int_0^t \; dt' \; \dot{z}_{t'} \nonumber\\
&=& z_0 + \frac{1}{\alpha} \int_0^t \; dt' \; \Big[F(z_{t'}) + \alpha u+ \xi_{t'}\Big]
\end{eqnarray}
is uniquely specified if the noise history is known.
Since we consider a Gaussian white noise, the probability to have the
noise history $\xi_t$ is given by  \cite{OM53}
\begin{equation}
P[\xi_{t}] \propto \exp \left[ - \frac{1}{4 k_{\rm B}T \alpha}
\int_0^t \; dt' \; \big(\xi_{t'}\big)^2 \right]
\end{equation}
According to Eq. (\ref{zm}), the probability of a trajectory $z_t$
starting from the fixed initial point $z_0$ is thus written as
\begin{equation}
P[z_{t}\vert z_0] \propto \exp \left[ - \frac{1}{4 k_{\rm B}T \alpha}
\int_0^t \; dt' \; \Big(\alpha \dot{z}- F(z) - \alpha u\Big)^2
\right]
\label{cond.prob}
\end{equation}
We remark that the corresponding joint probability is
obtained by multiplying the conditional probability (\ref{cond.prob})
with the stationary probability density (\ref{Pst}) of the initial
position as
\begin{equation}
P[z_{t}] \propto P[z_{t}\vert z_0] \, p_{\rm st}(z_0)
\label{joint.prob}
\end{equation}

To extract the heat dissipated along a trajectory, we
consider the probability of a given path over the
probability to observe the reversed path having also reversed the
sign of the driving $u$. The reversed path is thus defined by
$z_{t'}^{\rm R}= z_{t-t'}$ which implies $ \dot{z}_{t'}^{\rm R} = -
\dot{z}_{t-t'}$. Therefore, we find that
\begin{eqnarray}
\ln \frac{P_+[z_{t}\vert z_0]}{P_-[z_{t}^{\rm R}\vert z_0^{\rm R}]}
&=& -\frac{1}{4 k_{\rm B}T \alpha}  \int_0^t \; dt' \; \left[
\Big(\alpha \dot{z} - F(z) - \alpha u\Big)^2
- \Big(-\alpha\dot{z} - F(z) + \alpha u\Big)^2 \right] \nonumber \\
&=& \frac{1}{k_{\rm B}T}  \int_0^t \; dt' \; \Big(\dot{z} -
u\Big) \; F(z)  \nonumber \\
&=& \frac{1}{k_{\rm B}T} \left[ V(z_0) - V(z_t) - u \int_0^t \; dt'
\; F(z) \right] \nonumber \\
&=& \frac{Q_t}{k_{\rm B} T}
\label{ratio}
\end{eqnarray}
which is precisely the fluctuating heat (\ref{heat})
dissipated along the random path $z_t$
and expressed in the thermal unit $k_{\rm B}T$.
The detailed derivation of this
result is carried out
in Appendix \ref{detail}.
The dissipation can thus be related to time-symmetry breaking already
at the level of mesoscopic paths.
We notice that the mean value of the fluctuating quantity (\ref{ratio})
behaves as described by Eq. (\ref{mean}) so that the heat
dissipation rate vanishes on average at equilibrium and is positive otherwise.
Relations similar to Eq. (\ref{ratio}) are known
for Boltzmann's entropy production \cite{MN03},
for time-dependent systems \cite{S05,C99,KPV07},
and in the the context of Onsager-Machlup theory \cite{TC06}.
We emphasize that the reversal of $u$ is essential
to get the dissipated heat from the way
the path probabilities $P_+$ and $P_-$ differ.

The main difference between these path probabilities comes from the shift
between the mean values of the fluctuations under forward or backward
driving.  Indeed, the average position is equal to $\langle
z\rangle_+=u\tau_R$ under forward driving at
speed $+u$, and $\langle z\rangle_-=-u\tau_R$ under backward driving
at speed $-u$. The shift $2u\tau_R$ in the average positions implies
that a typical path of the forward time series falls,
after its time reversal, in the tail of the probability distribution $P_-$
of the backward time series.  Therefore, the probabilities $P_-$ of the time-reversed forward paths
in the {\it backward} time series are typically lower than the probabilities $P_+$
of the corresponding forward paths.  The above derivation (\ref{ratio})
shows that the dissipation can be obtained in terms of their ratio $P_+/P_-$.
We emphasize that this derivation holds for anharmonic potentials as
well as harmonic ones, so that the result is general in this respect.

In the stationary state, the mean entropy production (\ref{mean})
is given by averaging the dissipated heat (\ref{ratio}) over all
possible trajectories:
\begin{eqnarray}
  \frac{d_{\rm i}S}{dt} &=& \lim_{t\to\infty}
\frac{k_{\rm B}}{t} \left \langle \ln \frac{P_+[z_{t}\vert
z_0]}{P_-[z_{t}^{\rm R}\vert z_0^{\rm R}]} \right\rangle \nonumber\\
&=& \lim_{t\to\infty}
\frac{k_{\rm B}}{t} \left \langle \ln
\frac{P_+[z_{t}]}{P_-[z_{t}^{\rm R}]} \right\rangle \nonumber\\
&=& \lim_{t\to\infty}
\frac{k_{\rm B}}{t} \int {\cal D}z_{t}  \; P_+[z_{t}] \; \ln
\frac{P_+[z_{t}]}{P_-[z_{t}^{\rm R}]} \ \geq \ 0
\label{path.int}
\end{eqnarray}
The second equality in Eq. (\ref{path.int})
results from the fact that the terms at the
boundaries of the time interval $t$ are vanishing
for the statistical average in the long-time limit.
Equation (\ref{path.int}) relates the thermodynamic dissipation
to a so-called Kullback-Leibler distance \cite{KL51}
or relative entropy \cite{W78} between the forward process
and its time reversal. Such a connection between dissipation and
relative entropy has also been described elsewhere \cite{J06,KPV07,G04}.
Since the relative entropy is known to be always non negative,
the mean entropy production (\ref{path.int})
satisfies the second law of thermodynamics, as it should.
Accordingly, the mean entropy production vanishes at equilibrium
because of Eq. (\ref{mean}) or, equivalently, as the
consequence of detailed balance which holds at equilibrium
when the speed $u$ is zero (see Appendix \ref{detail}).

We point out that the heat dissipated along an
individual path given by Eq. (\ref{ratio}) is a fluctuating quantity
and may be either positive or negative.  We here face the paradox
raised by Maxwell that the dissipation is non-negative on average but
has an undetermined sign at the level of the individual stochastic
paths. The second law of thermodynamics holds for entropy production
defined after statistical averaging with the probability
distribution.  We remain with fluctuating mechanical quantities at
the level of individual mesoscopic paths or microscopic trajectories.


\section{Dynamical randomness and thermodynamic entropy production}
\label{NA}

The aim of this section is to present a method to characterize the property of
dynamical randomness in the time series and to show how this property
is related to the thermodynamic entropy production when the paths are compared
with their time reversals. In this way, we obtain a theoretical prediction on the
relationship between dynamical randomness
and thermodynamic entropy production.

\subsection{$(\varepsilon,\tau)$-entropies per unit time}

Dynamical randomness is the fundamental property of temporal disorder
in the time series.
The temporal disorder can be characterized by an entropy as well as
for the other kinds of disorder.
In the case of temporal disorder, we have an entropy per unit time,
which is the rate of production of
information by the random process, i.e., the minimum number of bits
(digits or nats) required to record the time series during one time
unit. For random processes which are continuous in time and in their
variable, the trajectories should be sampled with a resolution
$\varepsilon$ and with a sampling time $\tau$.  Therefore, the
entropy per unit time depends {\it a priori} on each one of them and
we talk about the $(\varepsilon,\tau)$-entropy per unit time. Such a
quantity has been introduced by Shannon as the rate of generating
information by continuous sources \cite{S48}.  The theory of this
quantity was developed under the names of $\varepsilon$-entropy
\cite{K56} and rate distortion function \cite{B71}.  More recently,
the problem of characterizing dynamical randomness has reappeared in
the study of chaotic dynamical systems. A numerical algorithm was
proposed by Grassberger, Procaccia and coworkers \cite{GP83,CP85} in
order to estimate the Kolmogorov-Sinai entropy per unit time.
Thereafter, it was shown that the same algorithm also applies to
stochastic processes, allowing us to compare the property of
dynamical randomness of different random processes \cite{GW93,G98}.
Moreover, these dynamic entropies were measured
for Brownian motion at equilibrium \cite{GBFSGDC98,BSFGGDC01}.
We here present the extension of this method
to out-of-equilibrium fluctuating systems.

Since we are interested in the probability of a given succession of states
obtained by sampling the signal $Z(t)$ at small time intervals
$\tau$, a multi-time random
variable is defined according to ${\mathbi Z} = [Z(t_0),
Z(t_0+\tau),\ldots, Z(t_0+n\tau-\tau)]$, which represents the
signal during the time period $t-t_0=n\tau$. For a stationary
process, the probability distribution does not depend on the initial time
$t_0$. From the point of view of probability theory, the process
is defined by the $n$-time joint probabilities
\begin{equation}
P_{s}({\mathbi z}; d{\mathbi z},\tau,n) = {\rm Pr} \{
{\mathbi z} < {\mathbi Z} < {\mathbi z}+d{\mathbi z};s \} =
p_{s}({\mathbi z}) \, d{\mathbi z}
\end{equation}
where $p_{s}({\mathbi z})$ denotes the probability density for
${\mathbi Z}$ to take the
values ${\mathbi z}=(z_0,z_1,\ldots,z_{n-1})$ at times $t_0+i\tau$
($i=0,1,2,...,n-1$)
for some nonequilibrium driving $s=u/\vert u\vert = \pm 1$.
Now, due to the continuous nature in time and in space of the
process, we will consider the probability $P_+({\mathbi Z}_m;
\varepsilon, \tau, n)$ for the trajectory to remain within a distance
$\varepsilon$ of some reference trajectory ${\mathbi Z}_m$, made of $n$
successive positions of the Brownian particle observed at time
intervals $\tau$ during the forward process. This reference
trajectory belongs to an ensemble
of $M$ reference trajectories $\{{\mathbi Z}_m\}_{m=1}^M$, allowing
us to take statistical averages.
These reference trajectories define the patterns, i.e., the recurrences of
which are searched for in the time series.

On the other hand, we can introduce the quantity $P_-({\mathbi
Z}^{\rm R}_m; \varepsilon,
\tau, n)$ which is the probability for a reversed trajectory of the
reversed process to remain within a distance $\varepsilon$ of the
reference trajectory ${\mathbi Z}_m$ (of the forward process) for $n$
successive
positions.

Suppose we have two realizations over a very long time interval $L\tau
\gg n\tau$ given by the time series $\{ z_{\pm}(k\tau)\}_{k=1}^{L}$,
respectively for the forward ($+$) and backward
($-$) processes. Within these
long time series, sequences of length $n$ are compared with each
other. We thus consider an ensemble set of $1 \ll M \ll L$ reference
sequences, which are all of
length $n$:
\begin{equation}
{\mathbi Z}_m = [Z(m\tau), \cdots, Z(m\tau+n\tau-\tau)] , \quad m \in
\{ m_1,\cdots,m_M \}
\end{equation}
These reference sequences are taken at equal time intervals in order
to sample the forward process according to its probability distribution $P_+$.
The distance between a reference sequence and another sequence of
length $n$ is defined by
\begin{equation}
{\rm dist}_n ({\mathbi Z}_m,{\mathbi Z}_j) = \max \{ \vert
Z(m\tau)-Z(j\tau) \vert, \cdots,\vert
Z(m\tau+n\tau-\tau)-Z(j\tau+n\tau-\tau) \vert \}
\label{dist}
\end{equation}
for $j=1,2,...,L'=L-n+1$.
The probability for this distance to be smaller than $\varepsilon$ is
then evaluated by
\begin{equation}
P_+({\mathbi Z}_m; \varepsilon, \tau, n) = \frac{1}{L'} \ {\rm Number}
\{ {\mathbi Z}_j : {\rm dist}_n({\mathbi Z}_m,{\mathbi Z}_j)\le
\varepsilon \}
\label{P+}
\end{equation}
The average of the logarithm of these probabilities over the
different reference sequences gives the block entropy
\begin{equation}
H(\varepsilon,\tau,n) = - \frac{1}{M} \sum_{m=1}^M \ln
P_+({\mathbi Z}_m; \varepsilon, \tau, n)
\label{pe}
\end{equation}
also called the mean pattern entropy.
The ($\varepsilon,\tau$)-entropy per unit time is then defined as the
rate of the
linear growth of the block entropy as the length
$n$ of the reference sequences increases \cite{GW93,GP83,CP85}:
\begin{equation}
h (\varepsilon,\tau) = \lim_{n \rightarrow \infty} \lim_{L',M
\rightarrow \infty} \frac{1}{\tau} \Big[ H(\varepsilon,\tau,n+1)
-H(\varepsilon,\tau,n) \Big]
\label{h}
\end{equation}

Similarly, the probability of a reversed trajectory in the reversed
process can be evaluated by
\begin{equation}
P_-({\mathbi Z}_m^{\rm R}; \varepsilon, \tau, n) = \frac{1}{L'} \ {\rm Number}
\{ \tilde{\mathbi Z}_j : {\rm dist}_n({\mathbi Z}_m^{\rm
R},\tilde{\mathbi Z}_j)\le
\varepsilon \}
\label{P-}
\end{equation}
where ${\mathbi Z}_m^{\rm R} = [Z(m\tau+n\tau-\tau),
\cdots,Z(m\tau)]$ is the time reversal of the reference path
${\mathbi Z}_m$ of the forward process, while
$\{\tilde{\mathbi Z}_j\}_{j=1}^{L'}$ are the paths of the reversed
process (with the opposite driving $-u$).
In similitude with Eqs. (\ref{pe}) and (\ref{h}), we may introduce
the time-reversed block entropy:
\begin{equation}
H^{\rm R} (\varepsilon,\tau,n) = - \frac{1}{M} \sum_{m=1}^M \ln
P_-({\mathbi Z}^{\rm R}_m; \varepsilon, \tau, n)
\label{rpe}
\end{equation}
and the time-reversed ($\varepsilon,\tau$)-entropy per unit time:
\begin{equation}
h^{\rm R} (\varepsilon,\tau) = \lim_{n \rightarrow \infty} \lim_{L',M
\rightarrow \infty} \frac{1}{\tau} \Big[ H^{\rm
R}(\varepsilon,\tau,n+1)-H^{\rm R}(\varepsilon,\tau,n) \Big]
\label{hR}
\end{equation}

We notice that the dynamical entropy (\ref{h}) gives the decay rate
of the probabilities to find paths within a distance
$\varepsilon$ from a typical path ${\mathbi Z}=(Z_0,Z_1,Z_2,...,Z_{n-1})$
with $Z_i=Z(t_0+i\tau)$:
\begin{equation}
P_+({\mathbi Z}; \varepsilon, \tau, n) \sim \exp[-n\tau h(\varepsilon, \tau)] \qquad (n\to\infty)
\label{prob}
\end{equation}
as the number $n$ of time intervals increases.
In the case of ergodic random processes, this property is known as
the Shannon-McMillan-Breiman theorem \cite{B65}.
The decay rate $h$ characterizes the temporal disorder, i.e., dynamical randomness,
in both deterministic dynamical systems and
stochastic processes \cite{GW93,ER85,GP83,CP85,G98}.
On the other hand, the time-reversed dynamical entropy (\ref{hR}) is the decay rate
of the probabilities of the time-reversed paths in the reversed process:
\begin{equation}
P_-({\mathbi Z}^{\rm R}; \varepsilon, \tau, n) \sim \exp[-n\tau h^{\rm R} (\varepsilon, \tau)]
\qquad (n\to\infty)
\label{prob.R}
\end{equation}
Since $h^{\rm R}$ is the decay rate of the
probability to find, in the backward process, the time-reversed path
corresponding to some typical path of the forward process,
the exponential $\exp(-h^{\rm R}\Delta t)$ evaluates the
amount of time-reversed paths among the typical paths (of duration $\Delta t$).
The time-reversed entropy per unit time $h^{\rm R}$ thus characterizes
the rareness of the time-reversed paths in the forward process.

The dynamical randomness of the stochastic process ruled by the
Langevin equation (\ref{zm}) can be characterized in terms of
its $(\varepsilon,\tau)$-entropy per unit time.  This latter is
calculated for the case of a harmonic potential in Appendix
\ref{App}. For small values of the spatial resolution $\varepsilon$,
we find that
\begin{equation}
h(\varepsilon,\tau) = \frac{1}{\tau} \ln\sqrt{\frac{\pi{\rm
e}D\tau_R}{2\varepsilon^2}\left(1-{\rm e}^{-2\tau/\tau_R}\right)}+
O(\varepsilon^2)
\label{hBP}
\end{equation}
with the diffusion coefficient of the Brownian particle:
\be
D= \frac{k_{\rm B}T}{\alpha}
\label{diff}
\ee
The $(\varepsilon,\tau)$-entropy per unit time increases as the
resolution $\varepsilon$ decreases, meaning that randomness is found
on smaller and smaller scales in typical trajectories of the Brownian
particle.
After having obtained the main features of the
$(\varepsilon,\tau)$-entropy per unit time,
we go on in the next subsection by comparing it with the
time-reversed $(\varepsilon,\tau)$-entropy per unit time,
establishing the connection with thermodynamics.

\subsection{Thermodynamic entropy production}

Under nonequilibrium conditions, detailed balance does not hold so that the
probabilities of the paths and their time reversals are different.
Similarly, the decay rates $h$ and $h^{\rm R}$ also differ.
Their difference can be calculated by evaluating the path integral
(\ref{path.int})
by discretizing the paths with the sampling time $\tau$ and
resolution $\varepsilon$
\begin{eqnarray}
  \frac{d_{\rm i}S}{dt} &=& \lim_{t\to\infty}
\frac{k_{\rm B}}{t} \int {\cal D}z_{t}  \; P_+[z_{t}] \; \ln
\frac{P_+[z_{t}]}{P_-[z_{t}^{\rm R}]}\nonumber\\
&=& \lim_{\varepsilon \rightarrow 0} \lim_{\tau  \rightarrow 0}
\lim_{n \rightarrow \infty} \lim_{L',M
\rightarrow \infty} \frac{k_{\rm B}}{n\tau} \frac{1}{M} \sum_{m=1}^M \ln
\frac{P_+({\mathbi Z}_m; \varepsilon, \tau, n)}{P_-({\mathbi Z}^{\rm
R}_m; \varepsilon, \tau, n)}
\end{eqnarray}
The statistical average is carried out over $M$ paths of the forward
process and thus corresponds to the average with the probability
$P_+[z_{t}]$. The logarithm of the ratio of probabilities can be
splitted into the difference between the logarithms of the
probabilities, leading to the difference of the block entropies
(\ref{rpe}) and (\ref{pe}).  The limit $n\to\infty$ of the
block entropies divided by $n$ can be evaluated from the differences
between the block entropy at $n+1$ and the one at $n$, whereupon the
$(\varepsilon,\tau)$-entropies per unit time (\ref{h}) and (\ref{hR})
appear. Finally, the mean entropy production
in the nonequilibrium steady state is given by the difference between
the time-reversed and direct $(\varepsilon,\tau)$-entropies per unit
time:
\begin{equation}
\frac{d_{\rm i}S}{dt} = \lim_{\varepsilon \rightarrow 0} \lim_{\tau
\rightarrow 0} \; k_{\rm B} \; \Big[ h^{\rm R} (\varepsilon,\tau) - h
(\varepsilon,\tau) \Big]
\label{ds}
\end{equation}

The difference between $h^{\rm R}$ and $h$ characterizes the time
asymmetry of the ensemble of typical paths effectively realized
during the forward process. Equation (\ref{ds}) shows
that this time asymmetry is related to the thermodynamic entropy production.
The entropy production is thus expressed as the difference of two
usually very large quantities which increase for $\varepsilon$ going
to zero \cite{GW93,G98}.
Out of equilibrium, their difference remains finite and gives the entropy
production. At equilibrium, the time-reserval symmetry $h^{\rm R}=h$
is restored so that their difference vanishes with the entropy production.
In the next two sections, the theoretical prediction (\ref{ds})
is tested experimentally.


\section{Driven Brownian motion}
\label{BM}

The first experimental system we have
investigated is a Brownian particle trapped by an optical tweezer, which
is composed by a large numerical aperture microscope objective ($\times
63$, $1.3$) and by an infrared laser beam with a wavelength of $980$~nm
and a power of $20$~mW on the focal plane. The trapped polystyrene
particle has a diameter of $2$~$\mu$m and is suspended in a $20\%$
glycerol-water solution. The particle is trapped at $20$~$\mu$m from the
bottom plate of the cell which is $200$~$\mu$m thick. The detection of
the particle position $z_t$ is done using a He-Ne laser and an
interferometric technique \cite{SGMS97}. This technique allows us to have
a resolution on the position of the particle of $10^{-11}$~m. In order to
apply a shear to the trapped particle, the cell is moved with a
feedback-controlled piezo actuator which insures a perfect linearity of
displacement \cite{note}.

The potential is harmonic: $V=kz^2/2$. The stiffness of
the potential is $k=9.62 \; 10^{-6} \; \rm{kg} \; \rm{s}^{-2}$. The
relaxation time is $\tau_R= \alpha/k = 3.05 \; 10^{-3} \; \rm{s}$, which
has been determined by measuring the decay rate of the autocorrelation
of $z_t$. The variable $z_t$ is acquired at the sampling frequency
$f=8192 \; {\rm Hz}$. The temperature is $T=298\; {\rm K}$.

The mean square displacement of the Brownian particle in the optical
trap is $\sigma=\sqrt{k_{\rm B}T/k}=20.7$ nm, while the diffusion
coefficient is $D=\sigma^2/\tau_R=1.4\times 10^{-13}$ m$^2$/s.
We notice that the relaxation time is longer than the sampling time
since their ratio is $f\tau_R=25$.

\begin{figure}[h]
\centering
\begin{tabular}{cc}
\rotatebox{0}{\scalebox{0.29}{\includegraphics{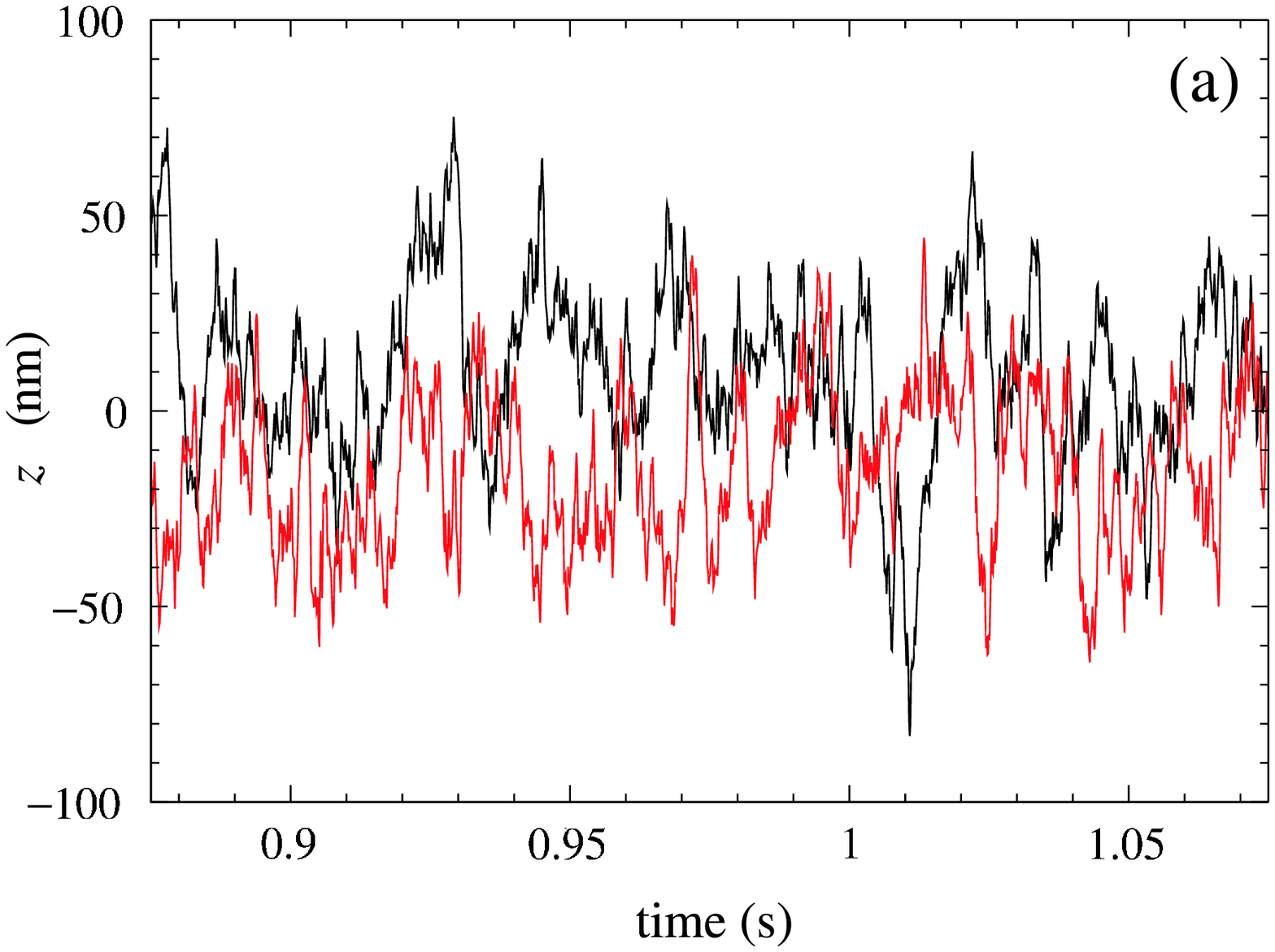}}} &
\rotatebox{0}{\scalebox{0.29}{\includegraphics{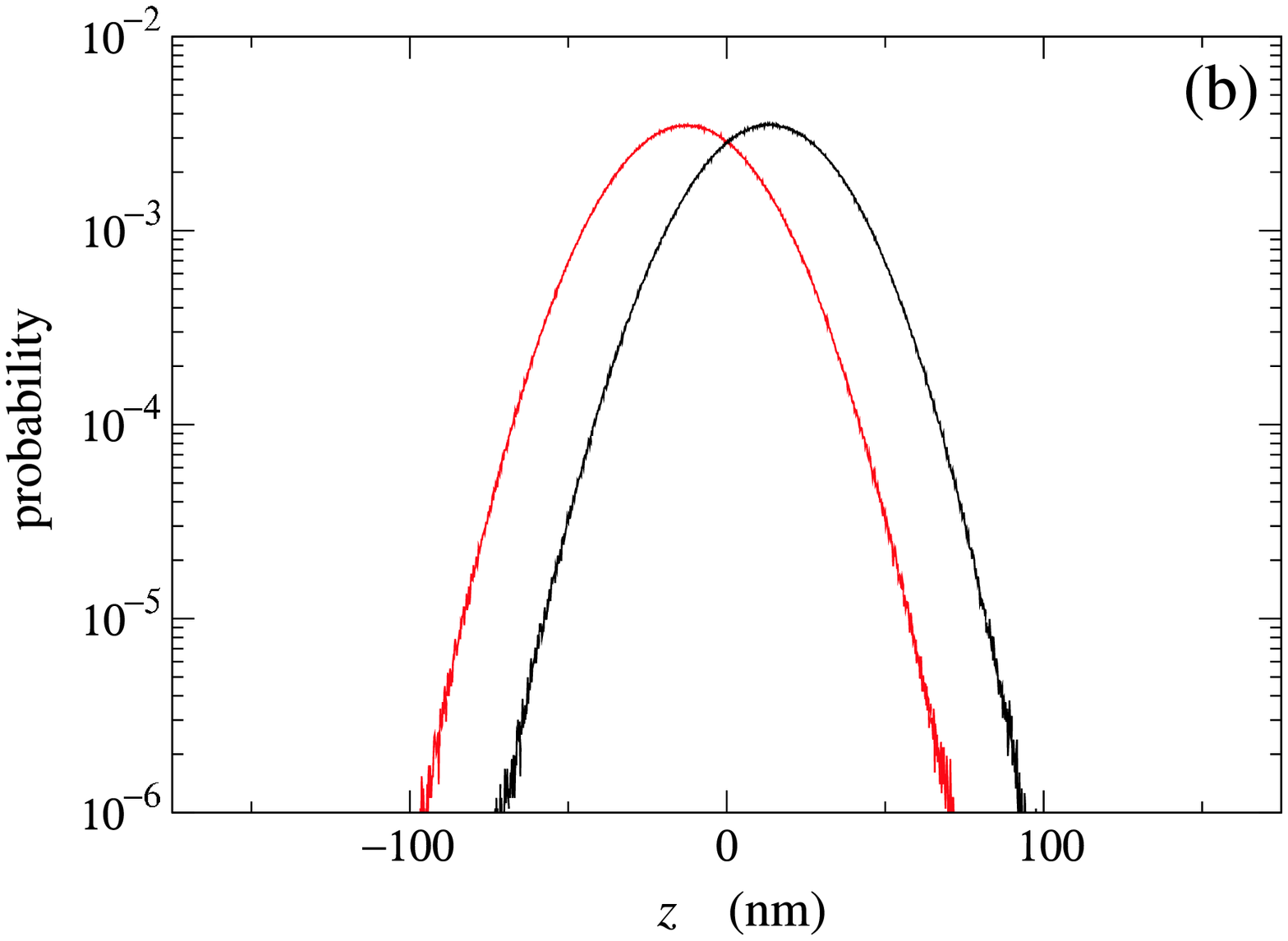}}} \\
\end{tabular}
\caption{(a) The time series of a typical path $z_t$ for the trapped Brownian
particle in the fluid moving
at the speed $u$ for the forward process (upper curve) and $-u$
for the reversed
process (lower curve) with $u=4.24 \times 10^{-6}$ m/s. (b) Gaussian
probability distributions of the forward and backward experiments.
The mean value is located at $\pm u\tau_R = \pm 12.9$ nm.}
\label{fig1}
\end{figure}

In order to test experimentally that entropy production is related to
the time asymmetry of dynamical randomness according to Eq. (\ref{ds}),
time series have been recorded for several values of $\vert u \vert$.
For each value, a pair of time series is generated,
one corresponding to the forward process
and the other to the reversed process,
having first discarded the transient evolution.
The time series contain up to $2 \times 10^7$ points each. Figure
\ref{fig1}a depicts examples of paths $z_t$ for the trapped Brownian
particle in the moving fluid. Figure \ref{fig1}b shows the
corresponding stationary distributions for the two time series.
They are Gaussian distributions shifted according to Eq. (\ref{Pst}).

\begin{figure}[h]
\centerline{\scalebox{0.7}{\includegraphics{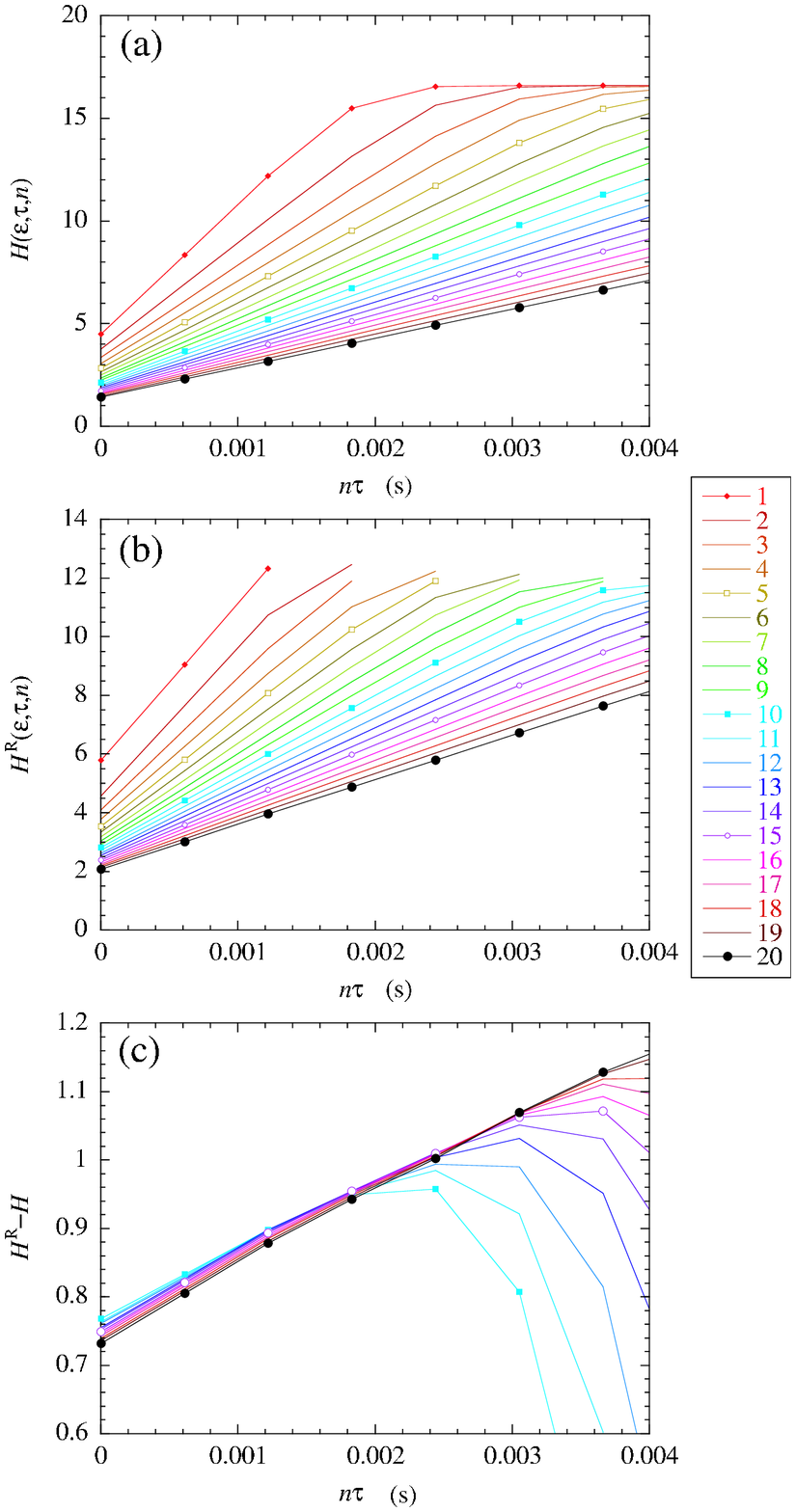}}}
\caption{(a) The block entropy (or mean pattern entropy) as a
function of time $n\tau$ for the trapped Brownian particle in the
fluid moving at the speed $u=4.24 \times 10^{-6}$ m/s. The time
interval $\tau$ is equal to the sampling time $\tau=1/f=1/8192$ s.
The different curves correspond to different values of
$\varepsilon = k \times 0.558 \ {\rm nm}$ with $k=1,\ldots,20$
given in the right-hand column. The distance used in this
calculation is defined by taking the maximum among the distances
$\vert Z(t)-Z_m(t)\vert$ for the times
$t=0,\tau,\ldots,(n-1)\tau$. The larger slopes correspond to the
smaller value of $\varepsilon$. The linear growth persists up to a
maximal value of the mean pattern entropy given by the total
length of the time series: $H_{\rm max} = \ln (1.4 \times 10^7)$.
(b) The mean reversed block entropy as a function of time
corresponding to (a). After a linear growth, $H^{\rm R}$ saturates
and falls down to zero (not shown) because of the finiteness of
the time series which limits the statistics. (c) Differences
between the backward and forward $(\varepsilon,\tau)$ dynamical
entropies in (b) and (a)
   for $\varepsilon$ between $5.6$-$11.2 \ {\rm nm}$ for the Brownian particle.
Straight lines are fitted to the first part of the curves and their
slopes give the
entropy production according to Eq. (\ref{ds}). }
\label{fig2}
\end{figure}

\begin{figure}[h]
\centerline{\scalebox{0.42}{\includegraphics{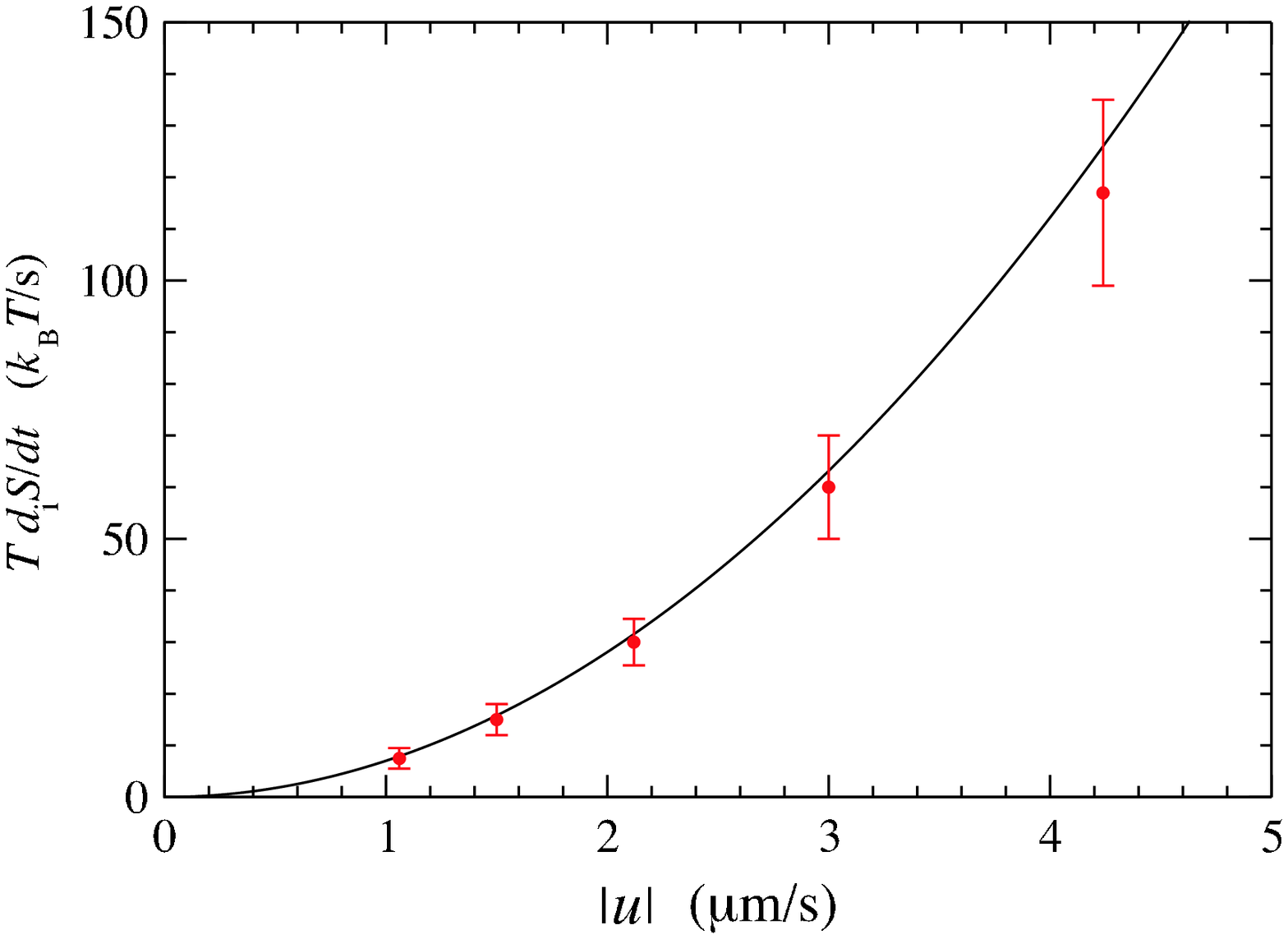}}}
\caption{Entropy production of the Brownian particle versus the
driving speed $u$. The solid line is the well-known rate of
dissipation given by Eq. (\ref{mean}). The dots depict the results
of Eq. (\ref{ds}) calculated with the differences between the
$(\varepsilon,\tau)$-entropies per unit time. The equilibrium
state is at zero speed $u=0$ where the entropy production
vanishes.} \label{fig3}
\end{figure}
\begin{figure}[h]
\begin{tabular}{cc}
\rotatebox{0}{\scalebox{0.5}{\includegraphics{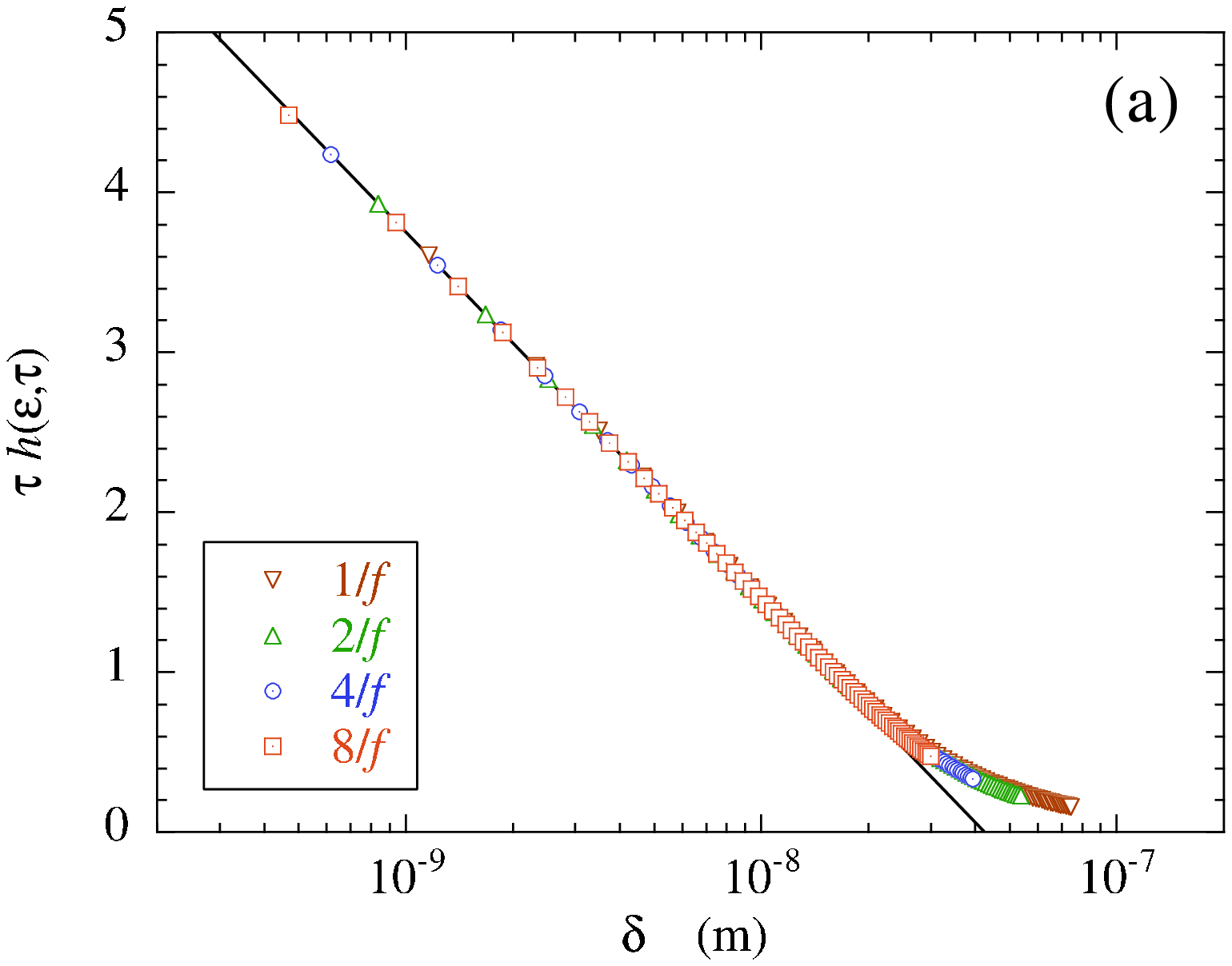}}} &
\rotatebox{0}{\scalebox{0.5}{\includegraphics{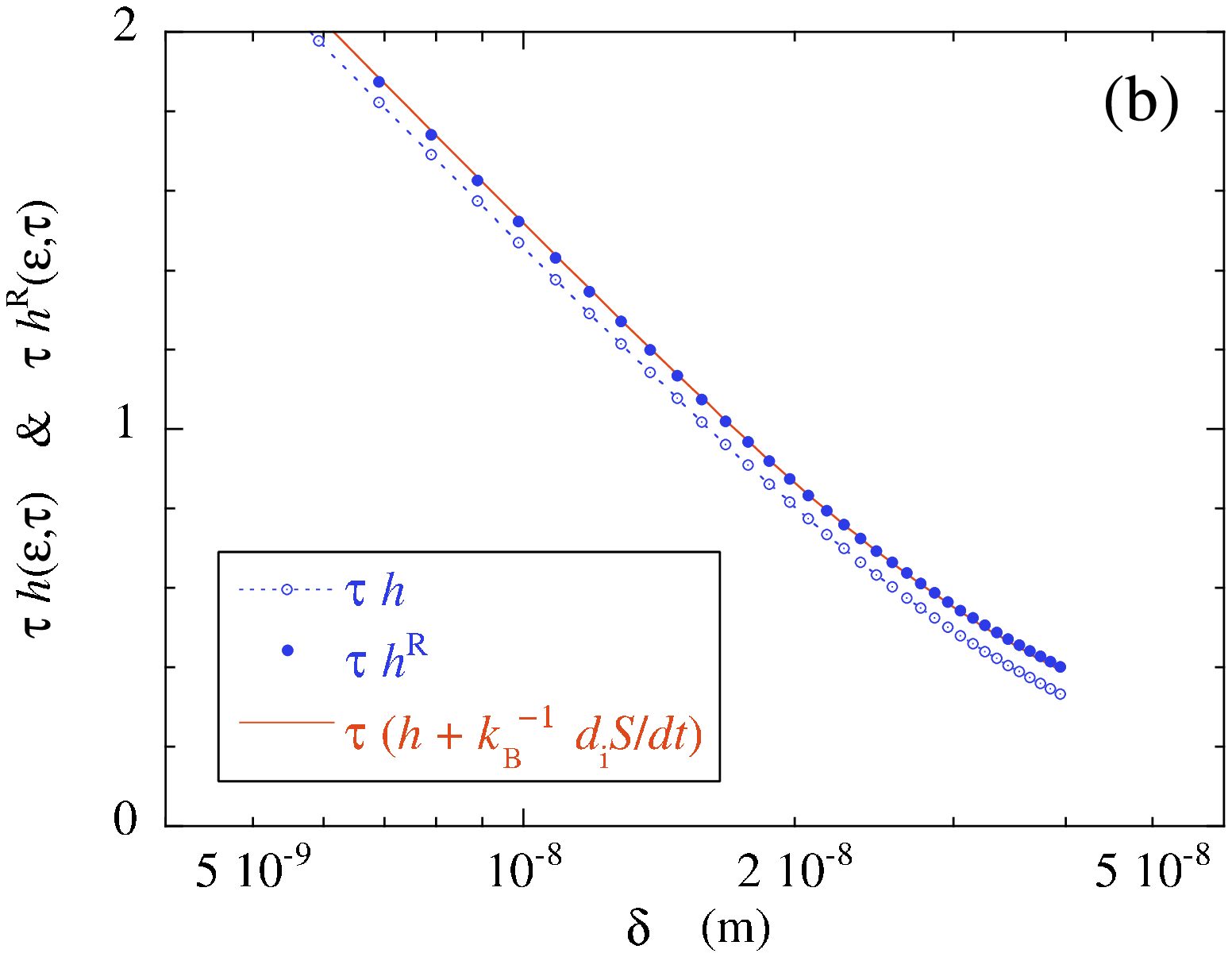}}} \\
\end{tabular}
\caption{(a) $(\varepsilon,\tau)$-entropy per unit time of the
Brownian particle scaled by the time interval $\tau$ as a function of
$\delta=\varepsilon/\sqrt{1-\exp(-2\tau/\tau_R)}$, for different
values of the time interval $\tau=1/f,2/f,4/f,8/f$, with the sampling
time $1/f=1/8192$~s. The dots are the results of the computation from
the time series for the speed $u= 4.24 \times 10^{-6}$~m/s. The solid
line depicts the expected behavior according to Eqs. (\ref{hBP}) and
(\ref{ds}). (b) Scaled reversed and direct
$(\varepsilon,\tau)$-entropies per unit time for $\tau=4/f$. The solid
line is the result expected from Eq. (\ref{ds}).}
\label{fig4}
\end{figure}
\begin{figure}[h]
\begin{tabular}{cc}
\rotatebox{0}{\scalebox{0.48}{\includegraphics{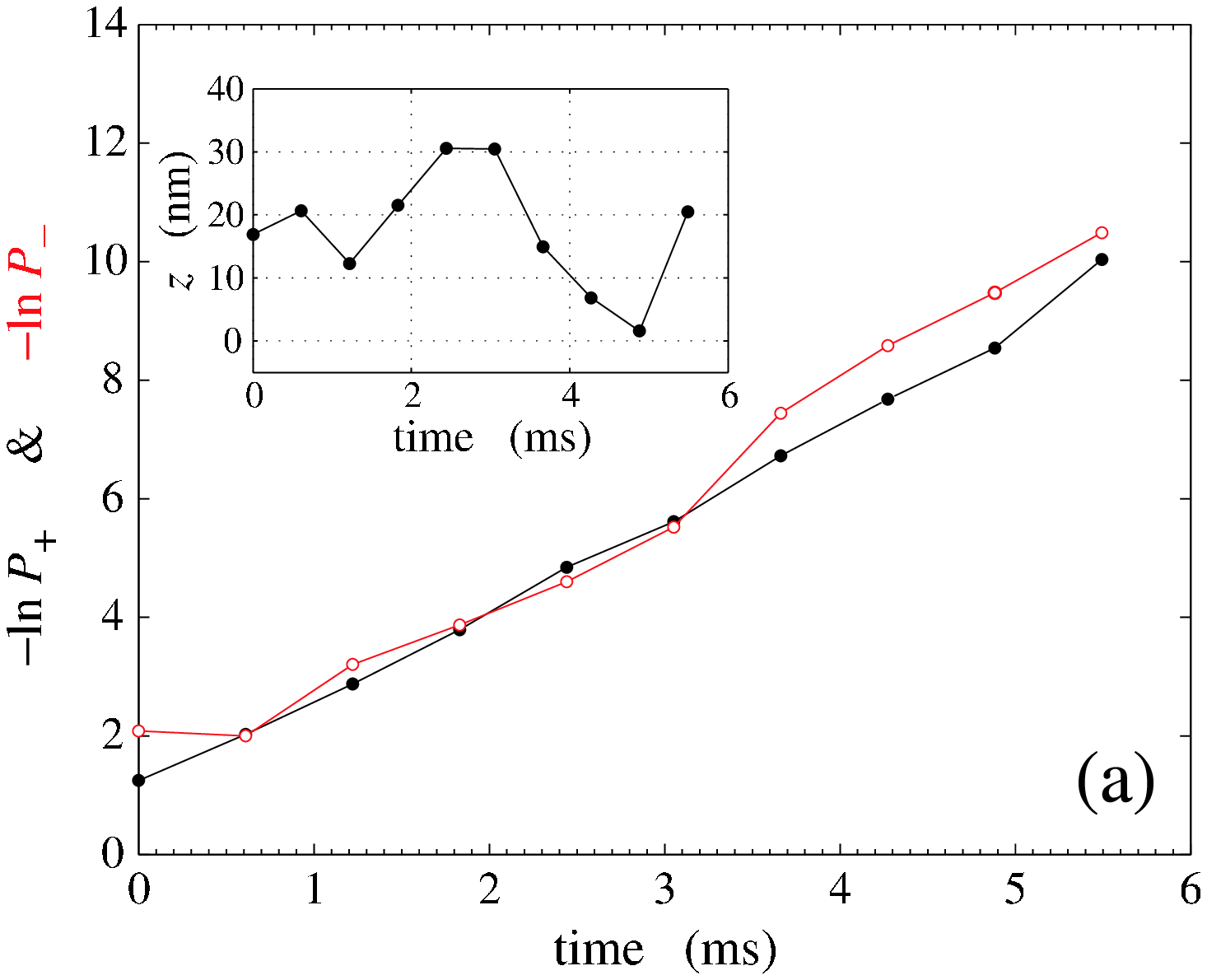}}} &
\rotatebox{0}{\scalebox{0.48}{\includegraphics{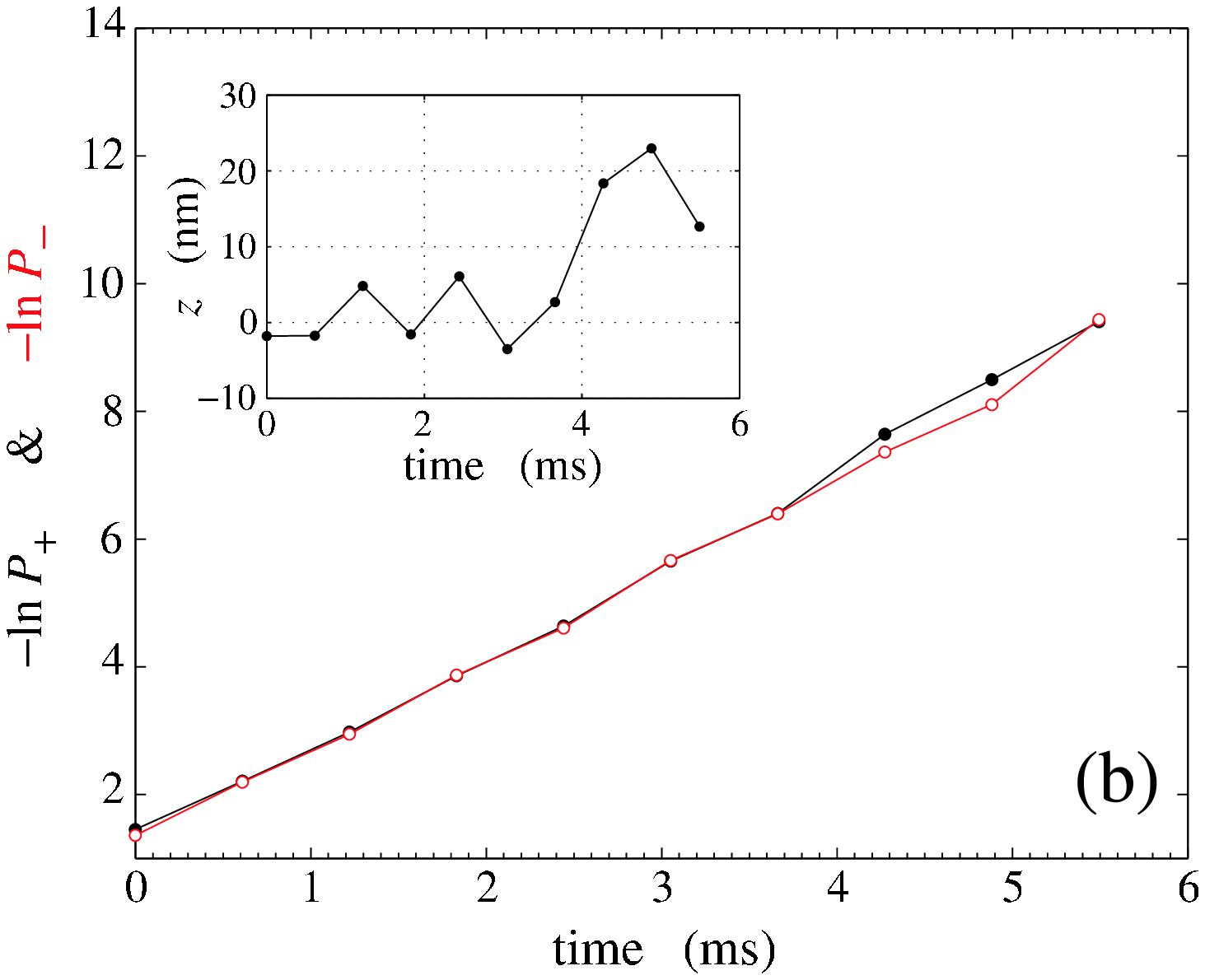}}} \\
\rotatebox{0}{\scalebox{0.29}{\includegraphics{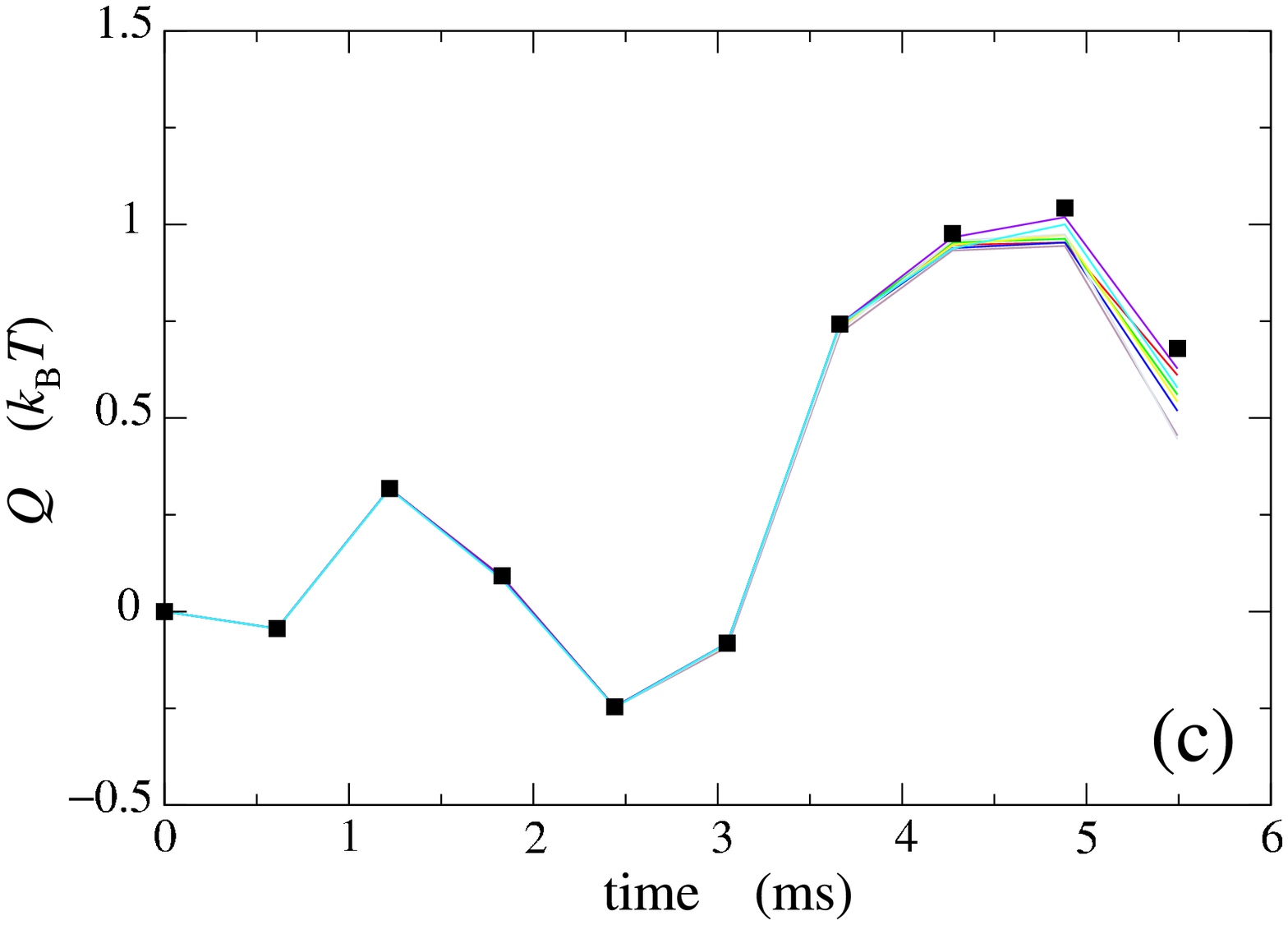}}} &
\rotatebox{0}{\scalebox{0.29}{\includegraphics{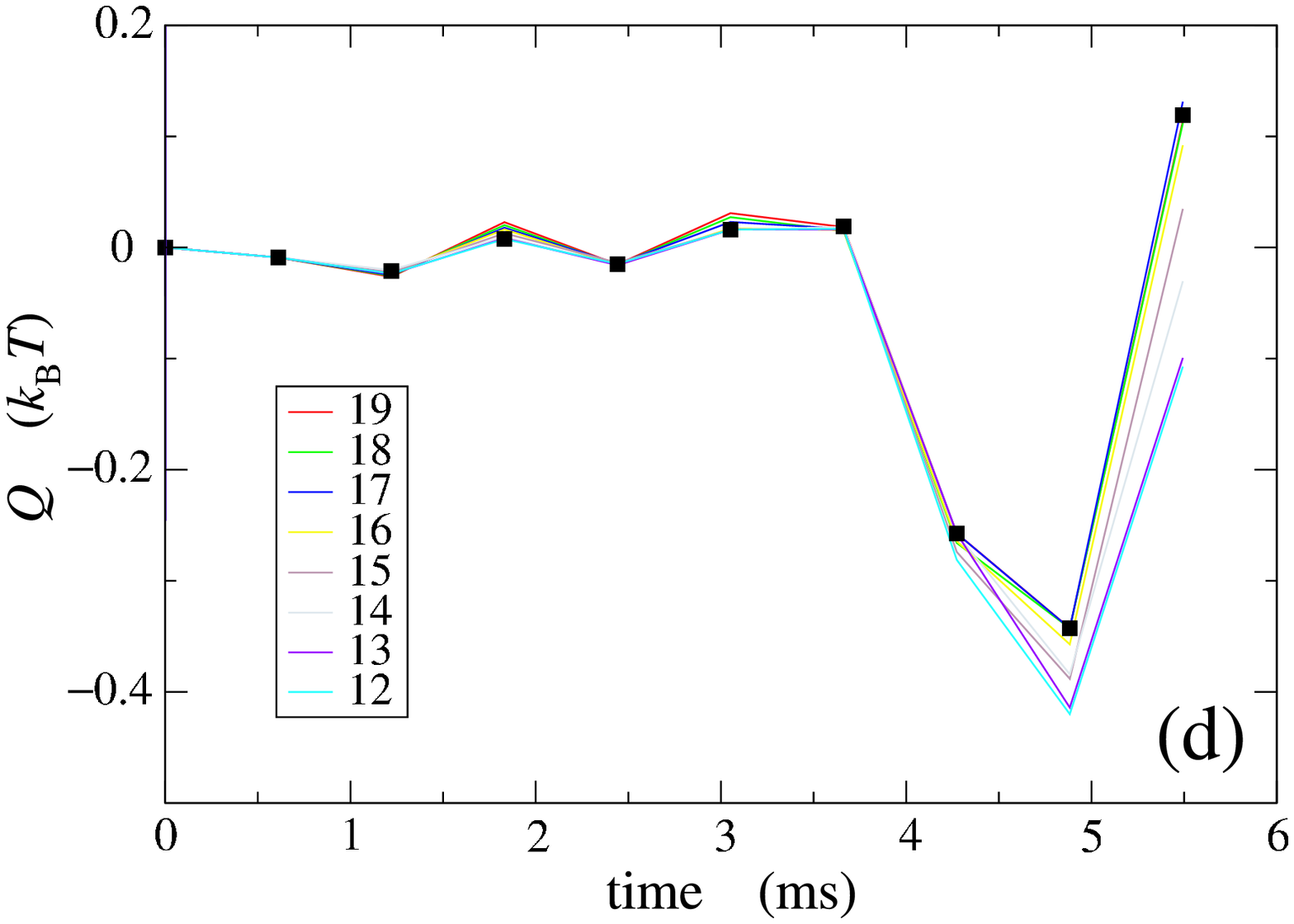}}} \\
\end{tabular}
\caption{Measure of the heat dissipated by the Brownian particle
along two randomly selected trajectories in the time series of Fig. \ref{fig1}.
The fluid speeds are $\pm u$ with $u=4.24 \times 10^{-6}$~m/s.
(a)-(b) Inset:  The two randomly selected trajectories.
The conditional probabilities $P_+$ and $P_-$
of the corresponding forward (filled circles) and the backward (open circles) paths
for $\varepsilon = 8.4\; {\rm nm}$, as evaluated by Eqs. (\ref{P+num}) and (\ref{P-num}).
These probabilities present an exponential decrease modulated by the fluctuations.
At time zero, the conditional probabilities are not defined
and, instead, we plotted at this time the stationary probabilities for indication.
(c)-(d) The dissipated heat given by the logarithm of the ratio of the forward
and backward probabilities according to Eq. (\ref{ratio}) for
different values of $\varepsilon = k \times 0.558 \ {\rm nm}$ with
$k=12,\ldots,19$
in the range $6.7$-$10.6\; {\rm nm}$.
They are compared with the values (squares) directly
calculated from Eq. (\ref{heat}). For small values of $\varepsilon$,
the agreement is quite good for short time
and are within experimental errors for larger time.} \label{fig5}
\end{figure}

The analysis of these time series is performed by calculating
the block entropy (\ref{pe}) versus the path duration $n\tau$,
and this for different values of $\varepsilon$. Figure \ref{fig2}a shows that
the block entropy increases linearly with the path duration $n\tau$
up to a maximum value fixed by the total length of the time series.
The time interval is taken equal to the sampling time: $\tau=1/f$.
The forward entropy per unit time $h(\varepsilon,\tau)$
is thus evaluated from the linear growth of the block entropy
(\ref{pe}) with the time $n\tau$.

Similarly, the time-reversed block entropy (\ref{rpe}) is computed
using the same reference sequences as for the forward block entropy,
reversing each one of them, and getting their probability of occurrence in the
backward time series.  The resulting time-reversed block entropy is depicted
in Fig. \ref{fig2}b versus $n\tau$ for different values of $\varepsilon$.
Here also, we observe that $H^{\rm R}$ grows linearly with the time $n\tau$
up to some maximum value due to the lack of statistics over long sequences
because the time series is limited.
Nevertheless, the linear growth is sufficiently extended
that the backward entropy per unit time $h^{\rm R}(\varepsilon,\tau)$
can be obtained from the slopes in Fig. \ref{fig2}b.

Figure \ref{fig2}c depicts the difference between the backward and
forward block entropies $H^{\rm R}$ and $H$ versus the time $n\tau$,
showing the time asymmetry due to the nonequilibrium constraint.
We notice that the differences  $H^{\rm R}-H$ are small compared with
the block entropies
themselves, meaning that dynamical randomness is large although the
time asymmetry is small.
Accordingly, the values $H^{\rm R}-H$ are more affected by the
experimental limitations than
the block entropies themselves.  In particular, the saturation due to
the total length of the time series affects the linearity of $H^{\rm
R}-H$ versus $n\tau$. However, we observe the expected independence
of the differences $H^{\rm R}-H$ on $\varepsilon$.  Indeed, the slope
which can be obtained from the differences $H^{\rm R}-H$ versus
$n\tau$ cluster around a common value (contrary to what happens for
$H$ and $H^{\rm R}$).  According to Eq. (\ref{ds}), the slope of
$H^{\rm R}-H$ versus $n\tau$ gives the thermodynamic entropy
production.

This prediction is indeed verified.  Figure \ref{fig3} compares the
difference $h^{\rm R}(\varepsilon,\tau)-h(\varepsilon,\tau)$ with the
thermodynamic entropy production given by the rate of dissipation
(\ref{mean}) as a function of the speed $u$ of the fluid. We
see the good agreement between both, which is the experimental
evidence that the thermodynamic entropy production is indeed related
to the time asymmetry of dynamical randomness. As expected, the
entropy production vanishes at equilibrium where $u=0$.

The dynamical randomness of the Langevin stochastic process can be
further analyzed by plotting the scaled entropy per unit time $\tau
h(\varepsilon,\tau)$ versus the scaled resolution $\delta \equiv
\varepsilon/\sqrt{1-\exp(-2\tau/\tau_R)}$ for different values of the
time interval $\tau$, as depicted in Fig. \ref{fig4}a. According to
Eq. (\ref{hBP}), the scaled entropy per unit time should behave as
$\tau h(\varepsilon,\tau) \simeq \ln(1/\delta)+C$ with some constant
$C$ in the limit $\delta\to 0$. Indeed, we verify in Fig. \ref{fig4}a
that, in the limit $\delta\to 0$, the scaled curves only depend
on the variable $\delta$ with the expected dependence $\ln(1/\delta)$.
For large values of $\delta$, the experimental curves deviate from the
logarithmic approximation (\ref{hBP}), since this latter is only
valid for $\delta\to 0$.  The calculation in Appendix \ref{App} shows
that we should expect corrections in powers of $\varepsilon^2$ to be
added to the approximation as $\ln(1/\delta)$.

In Fig. \ref{fig4}b, we depict the scaled direct and reversed
$(\varepsilon,\tau)$-entropies per unit time. We compare the behavior
of $h^{\rm R}$ with the behavior $\tau h^{\rm R} \simeq \tau \left( h
+ k_{\rm B}^{-1} d_{\rm i}S/dt\right)$ expected from the formula
(\ref{ds}). This figure shows that the direct and reversed
$(\varepsilon,\tau)$-entropies per unit time are quantities which are
large with respect to their difference due to the nonequilibrium
constraint.  This means that the positive entropy production is a
small effect on the top of a substantial dynamical
randomness.

Maxwell's demon vividly illustrates the paradox that the dissipated
heat is always positive at the macroscopic level although it may
take both signs if considered at the microscopic level of individual
stochastic trajectories. The resolution of Maxwell's paradox can be remarkably
demonstrated with the experimental data. Indeed, the heat dissipated
along an individual trajectories is given by Eq.
(\ref{heat}) and can be obtained by searching for recurrences in
the time series according to Eq. (\ref{ratio}). The conditional probabilities entering Eq. (\ref{ratio})
are evaluated in terms of the joint probabilities (\ref{P+}) and (\ref{P-}) according to
\begin{eqnarray}
P_+({\mathbi Z}; \varepsilon, \tau, n\vert Z_0)
&=& P_+({\mathbi Z}; \varepsilon, \tau, n)/P_+(Z_0; \varepsilon, \tau, 1)
\label{P+num}\\
P_-({\mathbi Z}^{\rm R}; \varepsilon, \tau, n\vert Z_{n-1})
&=& P_-({\mathbi Z}^{\rm R}; \varepsilon, \tau, n)/P_-(Z_{n-1}; \varepsilon, \tau, 1)
\label{P-num}
\end{eqnarray}
where we notice that the probabilities with $n=1$ are approximately equal
to the corresponding stationary probability density (\ref{Pst})
multiplied by the range $dz=2\varepsilon$.
The heat dissipated along two randomly selected paths are plotted in
Fig. \ref{fig5}.
We see the very good agreement between the values computed with Eq.
(\ref{heat}) using each path and Eq. (\ref{ratio}) using the
probabilities of recurrences in the time series.
We observe that, at the level of individual trajectories,
the heat exchanged between the particle and the surrounding fluid
can be positive or negative because of the molecular fluctuations.
It is only by averaging over the forward process that the dissipated heat takes
the positive value depicted in Fig. \ref{fig3}. Indeed, Fig.
\ref{fig3} is obtained after averaging over many reference paths as
those of Fig. \ref{fig5}. The positivity of the thermodynamic entropy
production results from this averaging, which solves Maxwell's
paradox.


\section{Electric noise in $RC$ circuits}
\label{RC}

The second system we have investigated is an $RC$ electric circuit
driven out of equilibrium by a current source which imposes the mean
current $I$ \cite{GC04}. The current fluctuates in the resistor because of the
intrinsic Nyquist thermal noise \cite{ZCC04}.  The $RC$ electric circuit
and the dragged Brownian particle, although physically different,
are known to be formally equivalent by the correspondence given in
Table \ref{table}.

The electric circuit is composed of a capacitor
with capacitance $C=278 \; \rm{pF}$
in parallel with a resistor of resistance
$R=9.22 \ {\rm M}\Omega$.  The relaxation time of the circuit is
$\tau_R = RC= 2.56 \times 10^{-3} \; \rm{s}$.
The charge $q_t$ going through  the resistor during the time interval
$t$ is acquired at the sampling frequency $f=8192 \; {\rm Hz}$.
The temperature is here also equal to $T=298\; {\rm K}$.

The mean square charge of the Nyquist thermal fluctuations is
$\sigma=\sqrt{k_{\rm B}TC}=6.7 \times 10^3 \, e$ where $e=1.602
\times 10^{-19}$ C is the electron charge.  The diffusion coefficient
is $D=\sigma^2/\tau_R=1.75\times 10^{10} \; e^2/{\rm s}$. The ratio
of the relaxation time to the sampling time is here equal to
$f\tau_R=21$.

\begin{figure}[h]
\centering
\begin{tabular}{cc}
\rotatebox{0}{\scalebox{0.29}{\includegraphics{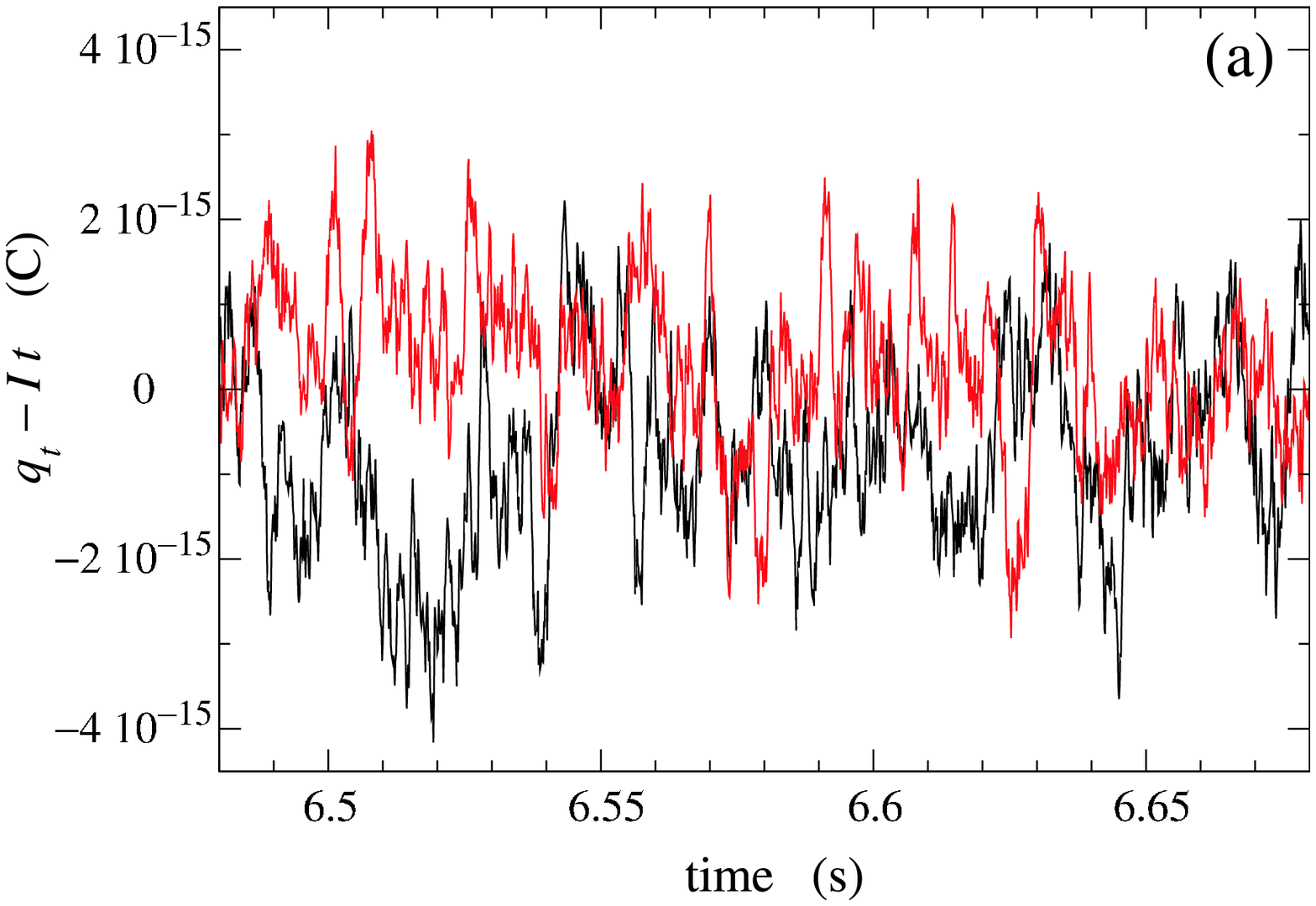}}} &
\rotatebox{0}{\scalebox{0.29}{\includegraphics{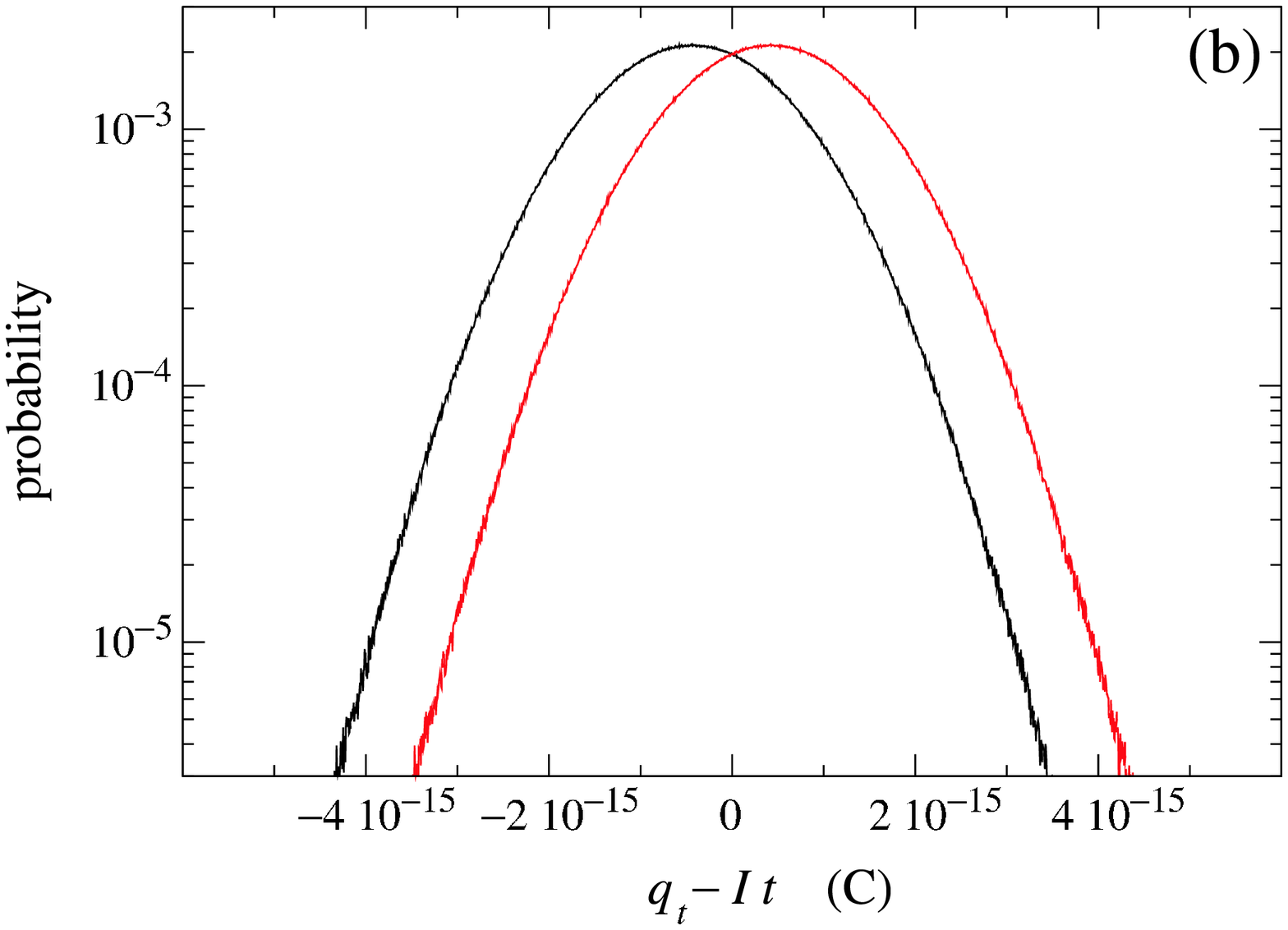}}} \\
\end{tabular}
\caption{(a) The time series of a typical path $q_t-It$ for the
Nyquist noise in the $RC$ electric circuit driven by the current $I$
(upper curve) and opposite current $-I$ (lower curve) with $I=1.67
\times 10^{-13}$~A. (b) Gaussian probability distributions of the
forward and backward experiments. The unit of the electric charge
$q_t-It$ is the Coulomb (C).}
\label{fig6}
\end{figure}

As for the first system, pairs of time series for opposite drivings
$\pm I$ are recorded.
Their length are $2 \times 10^7$ points each.
Figure \ref{fig6} depicts an example of a pair of such paths with the
corresponding probability distribution of the charge fluctuations.

\begin{figure}[h]
\begin{tabular}{cc}
\rotatebox{0}{\scalebox{0.5}{\includegraphics{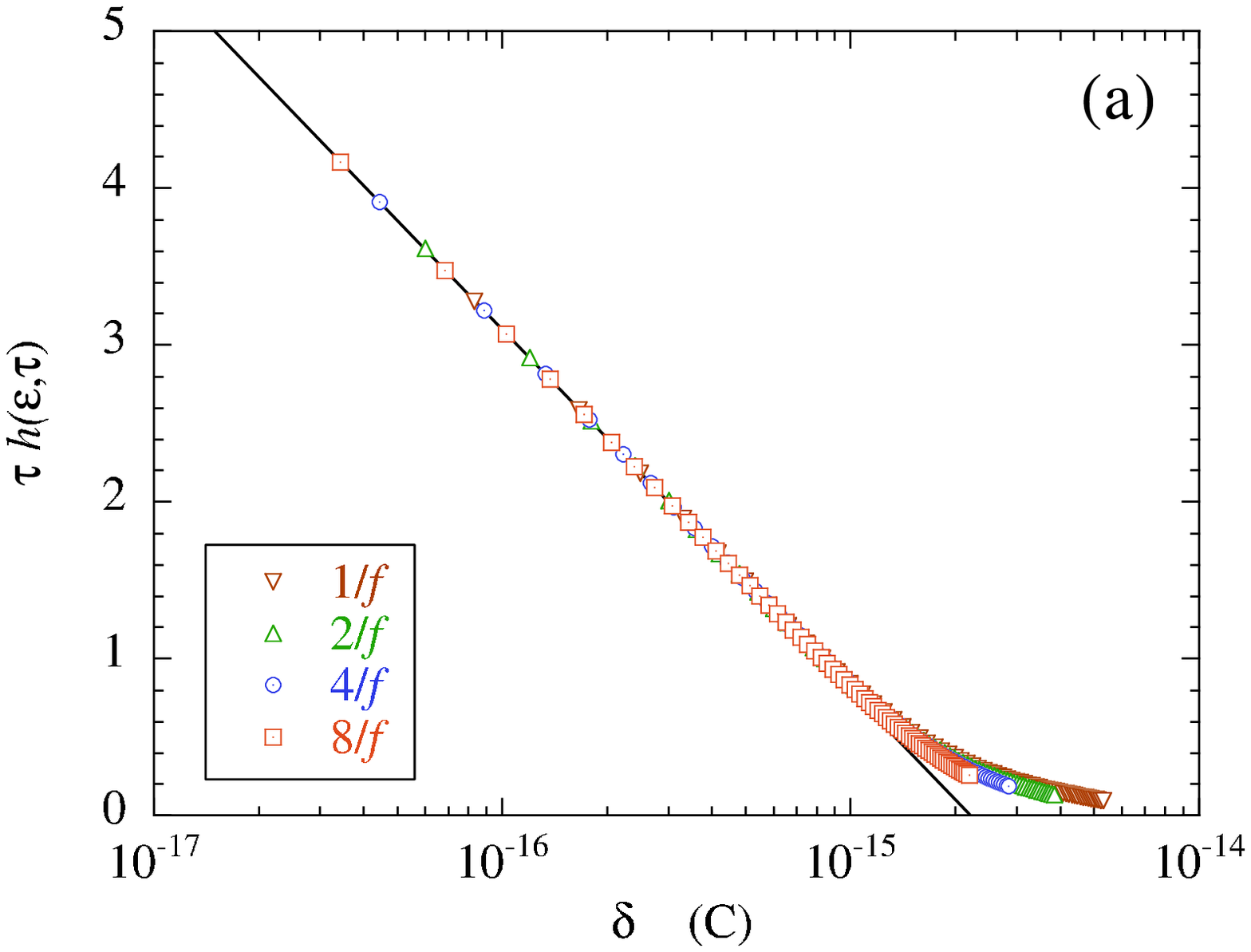}}} &
\rotatebox{0}{\scalebox{0.5}{\includegraphics{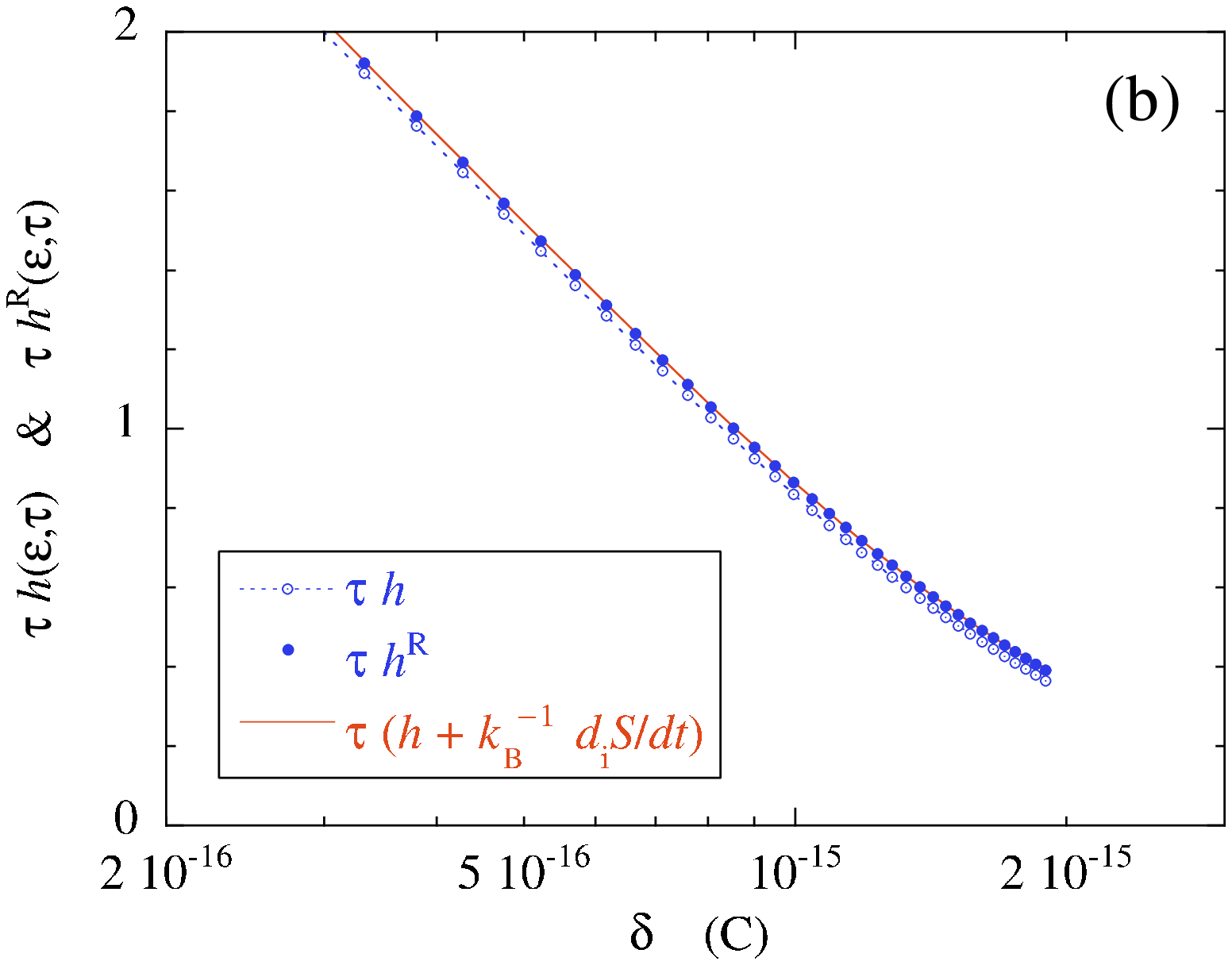}}} \\
\end{tabular}
\caption{(a) $(\varepsilon,\tau)$-entropy per unit time of the $RC$
electric circuit  scaled by the time interval $\tau$ as a function of
$\delta=\varepsilon/\sqrt{1-\exp(-2\tau/\tau_R)}$, for different
values of the time interval $\tau=1/f,2/f,4/f,8/f$, with the sampling
time $1/f=1/8192$ s. The dots are the results of the computation from
the time series for the current $I=1.67 \times 10^{-13}$~A. The solid
line depicts the expected behavior according to Eqs. (\ref{hBP})
and (\ref{ds}). (b) Scaled reversed and direct
$(\varepsilon,\tau)$-entropies per unit time for $\tau=4/f$. The solid
line is the result expected from Eq. (\ref{ds}). The unit of $\delta$
is the Coulomb (C).}
\label{fig7}
\end{figure}

The block entropies are here also calculated using Eqs. (\ref{pe})
and (\ref{rpe}) and the $(\varepsilon,\tau)$-entropies per unit time
are obtained from their linear growth as a function of the time
$n\tau$.
The scaled entropies per unit time $\tau h$ are depicted versus
$\delta$ in Fig. \ref{fig7}a.  Here again, the scaled entropy per
unit time is verified to depend only on $\delta$ for $\delta\to 0$,
as expected from the analytical calculation in Appendix \ref{App}.
In Fig. \ref{fig7}b, we compare the scaled reversed
$(\varepsilon,\tau)$-entropy per unit time to the behavior $\tau h^R
\simeq \tau \left( h  +k_{\rm B}^{-1} d_{\rm i}S/dt\right)$, expected
by our central result (\ref{ds}).

\begin{figure}[h]
\centerline{\scalebox{0.42}{\includegraphics{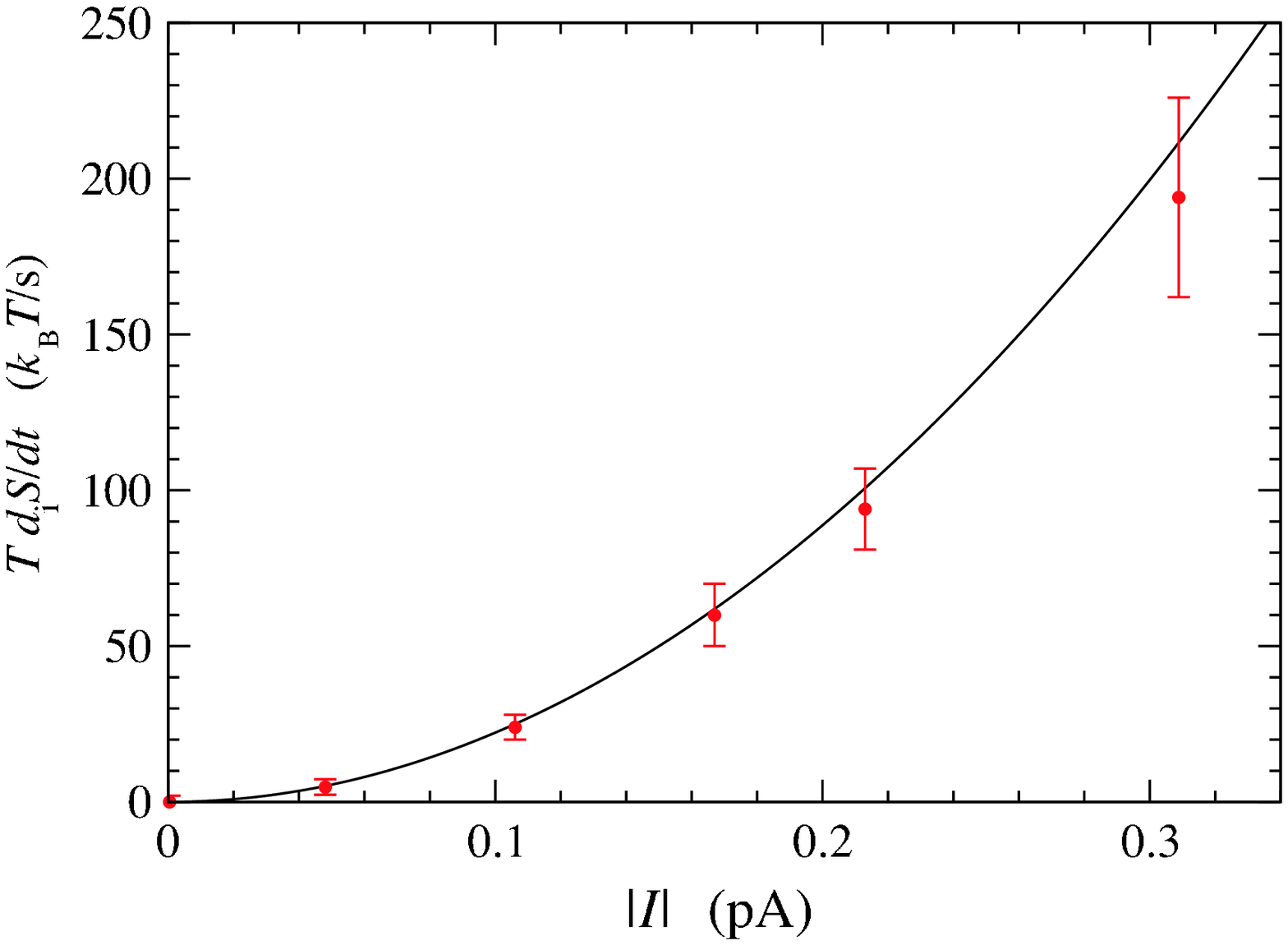}}}
\caption{Entropy production of the $RC$ electric circuit versus
the injected current $I$. The solid line is the Joule law,
$Td_{\rm i}S/dt=RI^2$. The dots are the results of Eq. (\ref{ds}).
The equilibrium state is at zero current, $I=0$, where the entropy
production is observed to vanish.} \label{fig8}
\end{figure}

The difference between the time-reversed and direct
$(\varepsilon,\tau)$-entropies per unit time is then compared with
the dissipation rate expected with Joule's law.  We observe the nice
agreement between both in Fig. \ref{fig8}, which confirms the
validity of Eq. (\ref{ds}). Here also, we observe that the entropy
production vanishes with the current at equilibrium.

\section{Discussion}
\label{discussion}

In this section, we discuss about the comparison between the present
results and other nonequilibrium relations.

The relation (\ref{ds}) expresses the entropy production as the difference between
the backward and forward $(\varepsilon,\tau)$-entropies per unit time.
The backward process is obtained by reversing the driving constraints,
which is also a characteristic feature of Crooks relation \cite{C99}.
However, Crooks relation is concerned with systems driven
by time-dependent external controls starting at equilibrium.
In constrast, our results apply to nonequilibrium steady states.
Another point is that Crooks relation deals with the fluctuations
of the work performed on the system,
while the present relation (\ref{ds}) gives the {\it mean}
value of the entropy production
and, this, in terms of path probabilities.
In this respect, the relation (\ref{ds}) is closer to a formula
recently obtained for the mean value of the dissipated work in systems
driven out of equilibrium by time-dependent external controls \cite{KPV07}.
This formula relates the mean value of the dissipated work to the logarithm
of the ratio between two phase-space probability densities associated with
the forward and backward processes, respectively. These phase-space probability
densities could in principle be expressed as path probabilities. Nevertheless,
these latter would be defined for systems driven over a finite time interval
starting from the equilibrium state, although the present equation (\ref{ds})
applies to nonequilibrium steady states
reached in the long-time limit.

We now compare our results to the extended fluctuation theorem,
which concerns nonequilibrium steady states \cite{ZC03,ZCC04,GC04}.
The extended fluctuation theorem is a symmetry relation of the
large-deviation properties of the fluctuating heat dissipation
(\ref{heat}) during a time interval $t$. The probability that this
fluctuating quantity takes the value
\be
\zeta \simeq \frac{1}{t}\frac{Q_t}{T}
\ee
decays exponentially with the rate
\be
G(\zeta) \equiv \lim_{t\to\infty} -\frac{1}{t} \ln {\rm Pr}\left\{
\zeta <  \frac{Q_t}{tT} < \zeta+d\zeta\right\}
\label{R}
\ee
This rate is a function of the value $\zeta$.  Since the variable
$\zeta$ can significantly deviate from the statistical average for
some large fluctuations, the rate (\ref{R}) is called a
large-deviation function.
The extended fluctuation theorem states
that the ratio of the probability of a positive fluctuation to the
probability of a negative one increases exponentially as $\exp(\zeta
t)$ in the long-time limit $t\to\infty$ and over a range of values of
$\zeta$, which is limited by its average $\langle\zeta\rangle$
\cite{ZC03,ZCC04,GC04}.  Taking the logarithm of the ratio and the
long-time limit, the extended fluctuation theorem can therefore be
expressed as the following symmetry relation for the decay rate
(\ref{R}):
\be
\zeta = k_{\rm B} \Big[ G(-\zeta)-G(\zeta)\Big] \qquad
\mbox{for}\quad  -\langle\zeta\rangle \leq \zeta \leq \langle\zeta\rangle
\ee
In this form, we notice the analogy with Eq. (\ref{ds}).  {\it A
priori}, the decay rate (\ref{R}) can be compared with the
$(\varepsilon,\tau)$-entropy per unit time, which is also a decay
rate.
However, the decay rate (\ref{R}) concerns the probability of all the
paths with dissipation $\zeta$
while the $(\varepsilon,\tau)$-entropy per unit time concerns the
probability of the paths within a distance $\varepsilon$ of some
reference typical paths. The $(\varepsilon,\tau)$-entropy per unit
time is therefore probing more deeply into the fluctuations down to
the microscopic dynamics. In principle, this latter should be reached
by zooming to the limit $\varepsilon\to 0$.

A closer comparison can be performed by considering the mean value of
the fluctuating quantity $\zeta$ which gives the thermodynamic
entropy production:
\be
\langle \zeta \rangle = \lim_{t\to\infty} \frac{1}{t}\frac{\langle
Q_t\rangle }{T} = \frac{d_{\rm i}S}{dt}
\ee
Since the decay rate (\ref{R}) vanishes at the mean value:
\be
G\big(\langle \zeta \rangle\big) = 0
\ee
we obtain the formula
\be
\langle \zeta \rangle = k_{\rm B} G\big(-\langle \zeta \rangle\big)
\ee
which can be quantitatively compared with Eq. (\ref{ds}) since both
give the thermodynamic entropy production. Although the time-reversed
entropy per unit time $h^{\rm R}(\varepsilon,\tau)$ is {\it a priori}
comparable with the decay rate $G(-\zeta)$, it turns
out that they are different and satisfy in general the inequality
$h^{\rm R}(\varepsilon,\tau) \geq G(-\langle\zeta\rangle)$ since the
entropy per unit time is always non negative $h(\varepsilon,\tau)
\geq 0$.  Moreover, $h(\varepsilon,\tau)$ is typically a large
positive quantity. The greater the dynamical randomness, the larger
the entropy per unit time $h(\varepsilon,\tau)$, as expected in the
limit where $\varepsilon$ goes to zero. This shows that the
$(\varepsilon,\tau)$-entropy per unit time probes finer scales in the
path space where the time asymmetry is tested.


\section{Conclusions}
\label{conclusions}

We have here presented detailed experimental results giving evidence
that the thermodynamic entropy production finds its orgin in the time
asymmetry of dynamical randomness in the nonequilibrium fluctuations
of two experimental systems.  The first is a Brownian particle trapped
by an optical tweezer in a fluid moving at constant speed $0\leq \vert u\vert <
4.3 \; \mu$m/s.  The second is the electric noise in an $RC$ circuit
driven by a constant source of current $0\leq \vert I \vert < 0.3$
pA.  In both systems, long time series are recorded, allowing us to
carry out the statistical analysis of their properties of dynamical
randomness.

The dynamical randomness of the fluctuations is characterized in
terms of $(\varepsilon,\tau)$-entropies per unit time, one for the
forward process and the other for the reversed process with opposite
driving.
These entropies per unit time measure the temporal disorder in the time series.
The fact that the stochastic processes is continuous implies that the
entropies per unit time depend on the resolution $\varepsilon$ and
the sampling time $\tau$.  The temporal disorder of the forward
process is thus characterized by the entropy per unit time
$h(\varepsilon,\tau)$, which is the mean decay rate of the path
probabilities.  On the other hand, the time asymmetry of the process
can be tested by evaluating the amount of time-reversed paths of the
forward process among the paths of the reversed process. This amount
is evaluated by the probabilities of the time-reversed forward paths
in the reversed process and its mean decay rate, which defines the
time-reversed entropy per unit time $h^{\rm R}(\varepsilon,\tau)$.
The time asymmetry in the process can be measured by the difference
$h^{\rm R}(\varepsilon,\tau)-h(\varepsilon,\tau)$.  At equilibrium
where detailed balance holds, we expect that the probability
distribution ruling the time evolution is symmetric under time
reversal so that this difference should vanish.  In contrast, out of
equilibrium, detailed balance is no longer satisfied and we expect
that the breaking of the time-reversal symmetry for the invariant
probability distribution of the nonequilibrium steady state.  In this
case, a non-vanishing difference is expected.

The analysis of the experimental data shows that the difference of
$(\varepsilon,\tau)$-entropies per unit time is indeed non vanishing.
Moreover, we have the remarkable result that the difference gives the
thermodynamic entropy production.  The agreement between the
difference and the thermodynamic entropy production is obtained for
the driven Brownian motion up to an entropy production of nearly $120
\, k_{\rm B}T/{\rm s}$.  For electric noise in the $RC$ circuit, the
agreement is obtained up to an entropy production of nearly $200 \,
k_{\rm B}T/{\rm s}$. These results provide strong evidence that the
thermodynamic entropy production arises from the breaking of
time-reversal symmetry of the dynamical randomness in
out-of-equilibrium systems.

\vskip 0.5 cm

{\bf Acknowledgments.} This research is financially supported by the
F.R.S.-FNRS Belgium, the ``Communaut\'e fran\c caise de Belgique'' (contract
``Actions de Recherche Concert\'ees'' No. 04/09-312), and by the French contract
ANR-05-BLAN-0105-01.

\appendix

\section{Path probabilities and dissipated heat}
\label{detail}

The detailed derivation of Eq. (\ref{ratio}) is here
presented for the case of the Langevin stochastic process ruled by
Eq. (\ref{zm}).  This stochastic process is
Markovian and described by a Green function $G(z,z_0;t)$ which is the conditional probability
that the particle moves to the position $z$ during the time interval
$t$ given that its initial position was $z_0$ \cite{C43}.
For small values $\tau$ of the time interval, this Green function reads
\begin{equation}
G(z,z_0;\tau) \simeq \sqrt{\frac{\beta\alpha}{4\pi\tau}} \, \exp
\left\{ - \frac{\beta\alpha}{4\tau}\left[z-z_0 - \tau \left(
\frac{F(z_0)}{\alpha}+u\right)\right]^2\right\}
\label{green.tau}
\end{equation}
If we discretize the time axis into small time intervals $\tau$, the path
probability (\ref{cond.prob})
becomes
\begin{equation}
P[z_{t}\vert z_0] \propto \prod_{i=1}^{n-1}
G(z_i,z_{i-1};\tau)
\end{equation}
with $(n-1)\tau=t$. The ratio (\ref{ratio}) of the direct and reversed path
probabilities is thus given by
\begin{equation}
\frac{P_+[z_{t}\vert z_0]}{P_-[z_{t}^{\rm R}\vert z_0^{\rm R}]}
\simeq \prod_{i=1}^{n-1}
\frac{G_+(z_i,z_{i-1};\tau)}{G_-(z_{n-i-1},z_{n-i};\tau)}
\end{equation}
where
the subscript is the sign of the fluid speed $u$. Inserting the
expression (\ref{green.tau}) for the Green functions, we get
\begin{eqnarray}
\ln \frac{P_+[z_{t}\vert z_0]}{P_-[z_{t}^{\rm R}\vert z_0^{\rm R}]}
&\simeq& -\frac{\beta\alpha}{4 \tau}  \sum_{i=1}^{n-1} \left\{ \left[z_i-z_{i-1} - \tau \left(
\frac{F(z_{i-1})}{\alpha}+u\right)\right]^2 -
\left[z_{n-i-1}-z_{n-i} - \tau \left(
\frac{F(z_{n-i})}{\alpha}-u\right)\right]^2 \right\} \nonumber\\
&=& \frac{\beta}{2} \sum_{i=1}^{n-1} \left[ (z_{i}-z_{i-1}) F(z_{i-1}) - (z_{n-i-1}-z_{n-i})F(z_{n-i})\right] -\frac{\beta u \tau}{2} \sum_{i=1}^{n-1} \left[ F(z_{i-1})+F(z_{n-i})\right]  +O(\tau)
\nonumber\\
&=& \beta \sum_{i=1}^{n-1} \frac{F(z_{i-1})+F(z_i)}{2}  (z_{i}-z_{i-1}) -\beta u \tau \sum_{i=1}^{n-1}  \frac{F(z_{i-1})+F(z_i)}{2} +O(\tau)
\end{eqnarray}
where we used the substitution $n-i \to i$ in the last terms of each sum.
In the continuous limit $n\rightarrow \infty$, $\tau \rightarrow 0$ with $(n-1)\tau=t$, the sums become integrals and we find
\begin{eqnarray}
\ln \frac{P_+[z_{t}\vert z_0]}{P_-[z_{t}^{\rm R}\vert z_0^{\rm R}]}
&=& \beta \left[ \int_{z_0}^{z_t} F(z)\, dz - u \int_0^t F(z_{t'})\, dt'  \right] \nonumber\\
&=& \beta  \left[ V(z_0)-V(z_t)  - u \int_0^t F(z_{t'})\, dt'  \right]
\label{ratiogen}
\end{eqnarray}
since $F(z)=-\partial_zV(z)$.  We thus obtain the expression given in Eq. (\ref{ratio}).

Now, we show that the detailed balance condition
\begin{equation}
\mbox{equilibrium:} \qquad P_+[z_{t}]=P_-[z_{t}^{\rm R}] \qquad (u=0)
\label{detailed.balance}
\end{equation}
holds in the equilibrium state when the speed of the fluid is set to zero.
Indeed, the equilibrium probability density of a Brownian particle
in a potential $V(z)$ is given by
\begin{equation}
p_{\rm eq} (z) = C \, \exp \left[ - \beta V(z) \right]
\label{Peq}
\end{equation}
with a normalization constant $C$. Since the joint probabilities
are related to the conditional ones by Eq. (\ref{joint.prob})
with $p_{\rm st}=p_{\rm eq}$ at equilibrium, we find
\begin{equation}
\ln \frac{P_+[z_{t}]}{P_-[z_{t}^{\rm R}]}\Big\vert_{\rm eq}
=\ln \frac{P_+[z_{t}\vert z_0]\, p_{\rm eq}(z_0)}{P_-[z_{t}^{\rm R}\vert z_0^{\rm R}]\, p_{\rm eq}(z_0^{\rm R})}\Big\vert_{\rm eq} = 0
\label{ratio.det.bal}
\end{equation}
The vanishing occurs at zero speed $u=0$ as a consequence of Eqs. (\ref{ratiogen}),
(\ref{Peq}), and $z_0^{\rm R}=z_t$. Hence, the detailed balance condition is satisfied at equilibrium.
Therefore, the last expression of Eq. (\ref{path.int}) vanishes at equilibrium with the entropy production, as expected.


\section{Dynamical randomness of the Langevin stochastic
process}
\label{App}

In this appendix, we evaluate the $(\varepsilon,\tau)$-entropy per
unit time
of the Grassberger-Procaccia algorithm for the Langevin
stochastic process of Eq. (\ref{zm}) with a harmonic trap
potential.  In this case, the Langevin process is an
Ornstein-Uhlenbeck stochastic process for the new variable
\begin{equation}
y\equiv z - u\tau_R
\end{equation}
The Langevin equation of this process
is
\begin{equation}
\frac{dy}{dt} = - \frac{y}{\tau_R} +
\frac{\xi_t}{\alpha}
\end{equation}
The probability density that the
continuous random variable $Y(t)$ takes the values ${\bf
y}=(y_0,y_1,...,y_{n-1})$ at the successive times
$0,\tau,2\tau,...,(n-1)\tau$ factorizes since the random process is
Markovian:
\begin{equation}
p(y_0,...,y_{n-1}) = G(y_{n-1},y_{n-2};\tau) \cdots G(y_1,y_0;\tau)
\, p_{\rm st}(y_0)
\label{path.prob.dens}
\end{equation}
with the Green function
\begin{equation}
G(y,y_0;t) = \frac{1}{\sqrt{2 \pi \sigma^2(1-{\rm e}^{-2 t/\tau_R})}}
\exp \left[ - \frac{(y - {\rm e}^{-t/\tau_R}y_0)^2}{2\sigma^2(1-{\rm
e}^{-2 t/\tau_R})} \right]
\label{green}
\end{equation}
with the variance
\begin{equation}
\sigma^2 = \frac{k_{\rm B}T}{k}
\label{variance}
\end{equation}
The stationary probability density is given by the Gaussian distribution
\begin{equation}
p_{\rm st}(y) = \lim_{t\to\infty} G(y,y_0;t) = \frac{1}{\sqrt{2 \pi \sigma^2}}
\exp \left( - \frac{y^2}{2\sigma^2} \right)
\label{p_st}
\end{equation}
Denoting by ${\bf y}^{\rm T}$ the transpose of the vector $\bf y$,
the joint probability density (\ref{path.prob.dens}) can be written
as the multivariate Gaussian distribution
\begin{equation}
p(y_0,...,y_{n-1}) = \frac{\exp\left(-\frac{1}{2} {\bf y}^{\rm
T}\cdot {\bf C}_n^{-1}\cdot{\bf
y}\right)}{(2\pi)^{\frac{n}{2}}(\det{\bf C}_n)^{\frac{1}{2}}}
\label{G.path.prob.dens}
\end{equation}
in terms of the correlation matrix
\bea
{\bf C}_n = \sigma^2
\left(\begin{array}{ccccc}
1 & r & r^2 & \ldots  &  r^{n-1}  \\
r & 1 & r & \ldots &  r^{n-2} \\
r^2 & r & 1 & \ldots  & r^{n-3}  \\
\vdots  & \vdots & \vdots  & \ddots &\vdots \\
r^{n-1} & r^{n-2} & r^{n-3} & \ldots & 1 \\
\end{array} \right)
\eea
with
\begin{equation}
r=\exp(-\tau/\tau_R)
\end{equation}
The inverse of the correlation matrix is given by
\bea
{\bf C}^{-1}_n = \frac{\sigma^{-2}}{(1-r^2)}
\left(\begin{array}{cccccc}
1 & -r & 0 & \ldots&0 & 0   \\
-r & 1+r^2 & -r & \ldots & 0 & 0 \\
0  & -r & 1+r^2& &0 &0 \\
\vdots & \vdots & \vdots & \ddots & \vdots& \vdots \\
0 & 0 & 0 & \ldots & 1+r^2& -r \\
0 & 0 & 0 & \ldots & -r& 1 \\
\end{array} \right)
\eea
and its determinant by
\bea
\det{\bf C}_n= \sigma^{2n}(1-r^2)^{n-1}
\label{det.exact}
\eea

The $(\varepsilon,\tau)$-entropy per unit time is defined by
\begin{equation}
h(\varepsilon,\tau) = \lim_{n \rightarrow \infty} -\frac{1}{n\tau}
\int_{-\infty}^{+\infty}dy_0 \cdots \int_{-\infty}^{+\infty}dy_{n-1}
\; p(y_0,...,y_{n-1}) \ln \int_{-\varepsilon}^{+\varepsilon}d\eta_0
\cdots \int_{-\varepsilon}^{+\varepsilon}d\eta_{n-1} \;
p(y_0+\eta_0,...,y_{n-1}+\eta_{n-1})
\label{hBP.int}
\end{equation}
where $\{y_j'=y_j+\eta_j\}_{j=0}^{n-1}$ represents the tube of
trajectories satisfying the conditions $\vert y_j'-y_j\vert <
\varepsilon$ with $j=0,1,...,n-1$, around the reference trajectory
sampled at the successive positions $\{y_j\}_{j=0}^{n-1}$. After
expanding in powers of the variables $\eta_j$ and evaluating the
integrals over $-\varepsilon<\eta_j<+\varepsilon$, the logarithm is
obtained as
\begin{equation}
\ln \int_{-\varepsilon}^{+\varepsilon}d\eta_0 \cdots
\int_{-\varepsilon}^{+\varepsilon}d\eta_{n-1} \;
p(y_0+\eta_0,...,y_{n-1}+\eta_{n-1}) =
\ln\frac{(2\varepsilon)^n}{(2\pi)^{\frac{n}{2}}(\det{\bf
C}_n)^{\frac{1}{2}}} -\frac{1}{2}\, {\bf y}^{\rm T}\cdot {\bf
C}_n^{-1}\cdot{\bf y} + O(\varepsilon^2)
\end{equation}
The integrals over $-\infty<y_j<+\infty$ can now be calculated to get
the result (\ref{hBP}) by using Eq. (\ref{det.exact}). We find
\begin{equation}
h(\varepsilon,\tau) = \frac{1}{\tau} \ln\sqrt{\frac{\pi{\rm
e}\sigma^2}{2\varepsilon^2}(1-r^2)}+ O(\varepsilon^2)
\label{h-App}
\end{equation}
Since the relaxation time is given by Eq. (\ref{t_R})
and the diffusion coefficient by Eq. (\ref{diff}), the variance of
the fluctuations can be rewritten as $\sigma^2=k_{\rm B}T/k=D\tau_R$.
Substituting in Eq. (\ref{h-App}), we obtain the
$(\varepsilon,\tau)$-entropy per unit time given by Eq.
(\ref{hBP}).
The above calculation shows that the
$(\varepsilon,\tau)$-entropy per unit time of the Ornstein-Uhlenbeck
process is of the form
\begin{equation}
h(\varepsilon,\tau) = \frac{1}{\tau} \;
\phi\left(\frac{D\tau_R}{\varepsilon^2},{\rm e}^{-\tau/\tau_R}\right)
\label{hBP-phi}
\end{equation}
with some function $\phi(s,r)$ of
$s=D\tau_R/\varepsilon^2=\sigma^2/\varepsilon^2$ and
$r=\exp(-\tau/\tau_R)$.

In the limit where the time interval $\tau$
is much smaller than the relaxation time $\tau_R$,
the only
dimensionless variable is the combination $D\tau/\varepsilon^2$.
In
this case, we recover the result that the
$(\varepsilon,\tau)$-entropy per unit time is given
by
\begin{equation}
h(\varepsilon,\tau) = \frac{1}{\tau} \;
\psi\left(\frac{2D\tau}{\varepsilon^2}\right) \qquad \mbox{for}\quad
\tau\ll\tau_R
\label{h-diff}
\end{equation}
This $(\varepsilon,\tau)$-entropy per unit time is characteristic of
pure diffusion without trap potential,
as previously shown
\cite{GW93,G98,GBFSGDC98,BSFGGDC01}.



\begin{thebibliography}{99}
\bibitem{note} In Ref. \cite{AGCGJP07}, the same experiment was described in the frame of coordinate $x=-z+ut$ where the fluid is at rest and the trap is moving.
\end{thebibliography}
\end{document}